\documentclass[showpacs,twocolumn,showkeys,amsmath,amssymb,pra]{revtex4-1}

\usepackage{bm}
\usepackage{bbm}
\usepackage{color}
\usepackage{graphicx}   
\newcommand{\rd}{{\mathrm d}}
\newcommand{\re}{{\mathrm e}}
\newcommand{\ri}{{\mathrm i}} 
\newcommand{\kB}{k_{\rm B}} 
\newcommand{\difrac}{\displaystyle\frac}
\newcommand{\wo}{\widetilde\omega}                

\begin{document}

\title[Parametrically driven harmonic oscillator]
      {Periodic thermodynamics of the parametrically driven harmonic 
       oscillator}

\author{Onno R. Diermann} 
\author{Helge Frerichs}
\author{Martin Holthaus}

\affiliation{Institut f\"ur Physik, Carl von Ossietzky Universit\"at, 
	D-26111 Oldenburg, Germany}
                  
\date{May 30, 2019}

\begin{abstract}
We determine the quasistationary distribution of Floquet-state occupation
probabilities for a parametrically driven harmonic oscillator coupled to a
thermal bath. Since the system exhibits detailed balance, and the canonical
representatives of its quasienergies are equidistant, these probabilities are 
given by a geometrical Boltzmann distribution, but its quasitemperature differs
from the actual temperature of the bath, being affected by the functional form 
of the latter's spectral density. We provide two examples of quasithermal 
engineering, {\em i.e.\/}, of deliberate manipulation of the quasistationary 
distribution by suitable design of the spectral density: We show that the 
driven system can effectively be made colder than the undriven one, and 
demonstrate that quasithermal instability can occur even when the system is 
mechanically stable.  
\end{abstract} 


\maketitle 


\section{Introduction}
\label{sec:1}

If a quantum system governed by a time-independent Hamiltonian possessing 
eigenstates with energies~$E_n$ is coupled to a thermal reservoir having the 
temperature~$T_{\rm bath}$, and is given time to equilibrate, after a while 
each state will be populated with probability proportional to its respective 
Boltzmann factor $\exp(-\beta E_n)$, where $\beta = 1/(\kB T_{\rm bath})$, 
with $\kB$ denoting the Boltzmann constant~\cite{LandauLifshitzV,Reif09,
Pathria11}. If a time-periodically driven quantum system possessing Floquet 
states with quasienergies~$\varepsilon_n$ is coupled to such a reservoir it 
will likewise approach a quasistationary state characterized by certain 
occupation probabilities~$p_n$ of its Floquet states, but such distributions 
of Floquet-state occupation probabilities are lacking the universality of 
their equilibrium counterpart. The determination of such distributions is 
a major task of what may be termed {\em periodic thermodynamics\/}, as it 
has been formulated in a programmatic manner by Kohn~\cite{Kohn01}, and 
approached constructively by Breuer {\em et al\/}.~\cite{BreuerEtAl00} So far, 
knowledge about such quasistationary Floquet-state distributions is quite 
limited, although some statements have been derived for special classes of 
systems~\cite{ShiraiEtAl15,Liu15}. 
Remarkably, a linearly driven harmonic oscillator interacting with a 
harmonic-oscillator heat bath retains the Boltzmann distribution of the bath, 
that is, the Floquet states of the linearly driven harmonic oscillator 
are occupied according to a Boltzmann distribution with the bath 
temperature~\cite{BreuerEtAl00,LangemeyerHolthaus14}. On the other hand, 
the Floquet substates of a spin~$s$ exposed to a circularly polarized driving 
force while being coupled to a heat bath develop a Boltzmann distribution with 
an effective quasi\-temperature which differs from the actual temperature 
of the bath~\cite{SchmidtEtAl19}. In the case of strongly driven anharmonic 
oscillators the quasi\-stationary Floquet-state distributions can be 
significantly influenced by phase-space structures of the corresponding 
classical system~\cite{BreuerEtAl00,KetzmerickWustmann10}. Intriguingly, the 
quasistationary distributions of driven-dissipative ideal Bose gases allow 
for Bose-Einstein condensation into multiple states~\cite{VorbergEtAl13,
SchnellEtAl18}. These somewhat random snapshots indicate that the subject 
of quasi\-stationary Floquet-state distributions merits systematic further 
investigation. 

In the present work we explore the periodic thermodynamics of a model system 
which is substantially richer than the linearly driven harmonic oscillator 
but still retains much of its analytical simplicity, namely, the harmonic 
oscillator with a periodically time-dependent spring function. The 
parametrically driven harmonic oscillator with an arbitrary time-dependence of 
its spring function has been the subject of several seminal studies, among 
others by Husimi~\cite{Husimi53}, Lewis and Riesenfeld~\cite{LewisRiesenfeld69},
and by Popov and Perelomov~\cite{PopovPerelomov69}, on the grounds of which 
it is a relatively straightforward exercise to derive the Floquet states 
which emerge when the spring function depends on time in a periodic 
manner~\cite{PopovPerelomov70,Combescure86,Brown91}. Nonetheless, we will sketch
the construction of the Floquet states in some detail in Sec.~\ref{sec:2} below,
as the precise knowledge of these states is necessary for specifying their 
coupling to the bath, and for computing the bath-induced transition rates. 
In Sec.~\ref{sec:3} we will then provide typical numerical examples for the 
variation of the system's quasienergies with the driving strength, focusing on 
the archetypal case for which the classical equation of motion reduces to 
the well-known Mathieu equation. The central Sec.~\ref{sec:4} outlines the 
calculation of corresponding quasistationary distributions; this calculation 
is greatly facilitated by the fact that the system exhibits detailed balance.
We also consider the energy dissipation rate pertaining to the nonequilibrium
steady state for various spectral densities of the bath. One of the noteworthy 
benefits of this elemental model of periodic thermodynamics lies in the fact 
that it also provides a particularly transparent, analytical access to the 
question how the quasistationary distribution is affected if the spectral 
density is modified or, phrased the other way round, how the spectral density 
has to be designed in order to obtain quasistationary distributions with 
certain desired properties; this option of {\em quasithermal engineering\/} 
is discussed in the concluding Sec.~\ref{sec:5}.

\section{Floquet states of the parametrically driven harmonic oscillator}
\label{sec:2}

Consider a quantum particle of mass~$M$ moving in a one-dimensional 
harmonic-oscillator potential with time-dependent spring function $k(t)$,
as described in the position representation by the Hamiltonian
\begin{equation}
	H_0(t) = \frac{p^2}{2M} + \frac{1}{2} k(t) x^2 \; .
\label{eq:DHO}
\end{equation}
Later on we will focus on spring functions which depend {\em periodically\/}
on time~$t$, but at this point we still admit an arbitrary variation of $k$ 
with~$t$, requiring only the existence of the solutions to the corresponding
classical equations of motion; the significance of this requirement will
soon become obvious. For solving the time-dependent Schr\"odinger equation  
\begin{equation}
	\left( H_0(t) - \ri\hbar\frac{\partial}{\partial t} \right) 
	\psi(x,t) = 0
\label{eq:SGL}
\end{equation} 
we follow a strategy devised by Brown~\cite{Brown91}, and apply a sequence 
of two unitary transformations to Eq.~(\ref{eq:SGL}) which bring the
Hamiltonian~(\ref{eq:DHO}) into a more convenient form. Intending to replace 
the time-dependent potential $k(t) x^2/2$ by a more tractable one, we perform 
a first transformation which is implemented by~\cite{Brown91}    
\begin{equation}
	U_1 = \exp\big(-\ri\eta(t) x^2/\hbar\big) \; ,
\label{eq:UT1}
\end{equation}
where the function $\eta(t)$ will be suitably specified below. Using the 
familiar Lie expansion formula~\cite{Miller72} 
\begin{eqnarray}
	\re^A B \re^{-A} & = & B + [A,B] + \frac{1}{2} \Big[ A, [A,B] \Big]
	+ \ldots 
\nonumber \\ & & 	
	+ \frac{1}{n!} \Big[ A, \ldots [A,B] \Big]_{(n)} + \ldots
\label{eq:BCH}
\end{eqnarray}	
for operators $A$ and $B$, which reduces to
\begin{equation}
	\re^A B \re^{-A} = B + [A,B]
\end{equation}
if $\big[ A, [A,B] \big] = 0$, one finds 
\begin{eqnarray}
	U_1 x \, U_1^\dagger & = & x
\nonumber \\	
	U_1 p \, U_1^\dagger & = & p + 2\eta(t) x
\nonumber \\
	U_1 \big( -\ri\hbar \partial/\partial t \big) U_1^\dagger & = &
	 -\ri\hbar \partial/\partial t +\dot{\eta}(t) x^2 \; .
\end{eqnarray}
Taken together, these relations yield
\begin{eqnarray}
	& & 
	U_1 \left( \frac{p^2}{2M} + \frac{1}{2} k(t) x^2 
	- \ri\hbar \frac{\partial}{\partial t} \right) U_1^\dagger
\nonumber \\ 	& = &
	\frac{p^2}{2M} + \left[ \frac{2}{M} \eta^2(t) + \frac{1}{2} k(t)
	+ \dot{\eta}(t) \right] x^2 	
	- \ri\hbar \frac{\partial}{\partial t}
\nonumber \\ 	& + &	
	\frac{\eta(t)}{M} \big( p x + x p \big) \; .
\label{eq:TR1}
\end{eqnarray}	
For constructing the required counterterm to the coefficient $k(t)/2$ of 
$x^2$ appearing in the square brackets here, we resort to a solution 
$\xi(t) \equiv \xi$ to the classical equation of motion  
\begin{equation}
	M \ddot \xi  + k(t) \xi = 0 \; ;
\label{eq:EOM}	
\end{equation}
for ease of notation, the time-dependence of $\xi$ will not be indicated 
explicitly. If we demand that $\xi$ be {\em complex\/}, its conjugate $\xi^*$
is a second, linearly independent solution to Eq.~(\ref{eq:EOM}). It is easy 
to show that the Wronskian of these two solutions is time-independent, and 
purely imaginary, so that we may write
\begin{equation}
	\left| 	\begin{array}{cc}
		\dot{\xi} & \dot{\xi}^* \\
		     \xi  &      \xi^*
		\end{array} \right|
	= 2 \ri \Omega \; ,		        
\label{eq:WRO}
\end{equation}
where
\begin{equation}
	\Omega = {\rm Im} (\dot{\xi} \xi^*) \; ;
\end{equation}	
without loss of generality (that is, interchanging $\xi$ and $\xi^*$ if 
necessary), we may stipulate $\Omega > 0$. Given these classical solutions 
$\xi$ and $\xi^*$, we now set~\cite{Brown91}
\begin{equation}
	\eta(t) = \frac{M}{4} \left( \frac{\dot{\xi}}{\xi} 
	+ \frac{\dot{\xi}^*}{\xi^*} \right)
	= \frac{M}{4} \frac{\rd}{\rd t} \ln | \xi |^2 \; ,	
\label{eq:FET}
\end{equation}
providing
\begin{eqnarray}
	\dot \eta(t) & = & 
	\frac{M}{4} \left( 
	\frac{\ddot\xi}{\xi} - \frac{\dot\xi^2}{\xi^2}
	+ \frac{\ddot{\xi}^*}{\xi^*} - \frac{\dot\xi^{*2}}{\xi^{*2}}
	\right)
\nonumber \\	& = &
	\frac{M}{4}\left( - \frac{2 k(t)}{M} 
	- \frac{\dot\xi^2 \xi^{*2} + \dot\xi^{*2}\xi^2}{|\xi|^4} \right) \; .		
\end{eqnarray}
This is how the counterterm comes into play, but in view of Eq.~(\ref{eq:TR1})
we also need to account for a further contribution to the transformed 
quadratic potential, given by
\begin{equation}
	\frac{2}{M} \eta^2(t) = \frac{M}{8} \frac{1}{|\xi|^4}
	\left( \dot\xi^2 \xi^{*2} + 2\xi \, \dot\xi \, \xi^* \, \dot\xi^*
	+ \xi^2 \dot\xi^{*2} \right) \; .
\end{equation}			
Adding up, and making use of the Wronskian~(\ref{eq:WRO}), one finds
\begin{equation}
	\frac{2}{M}\eta^2(t) + \frac{1}{2} k(t) + \dot\eta(t)
	= \frac{1}{2} M \Omega^2 \frac{1}{| \xi |^4} \; .
\end{equation}	
At this point, it may be appropriate to point out that the construction still
leaves us with some indeterminacy: The classical equation~(\ref{eq:EOM}) is 
homogeneous, so that $\xi$ may be taken to be dimensionless, and we are free 
to multiply any solution $\xi$ by an arbitrary constant. Thus, $\Omega$ is 
not well defined, but $\Omega/|\xi|^2$ is. Note also that this multiplicative 
freedom does not affect the function $\eta(t)$, as a consequence of its 
definition~(\ref{eq:FET}), implying that the transformation~(\ref{eq:TR1}) 
indeed is unique.

At a first glance, it seems that this transformation~(\ref{eq:TR1}) with the 
particular choice~(\ref{eq:FET}) for the function~$\eta(t)$ has not brought 
us any further. On the contrary: Effectively, the spring function~$k(t)$ 
with its known time-dependence has been replaced by $M\Omega^2/|\xi|^4$, 
the time-dependence of which still needs to be determined by solving the 
classical Eq.~(\ref{eq:EOM}). However, the actual progress achieved by the 
operation~(\ref{eq:UT1}) stems from the last term on the right-hand side of 
Eq.~(\ref{eq:TR1}). Namely, one evidently has    
\begin{eqnarray}
	\left[ \frac{\ri}{\hbar} \frac{xp + px}{2} , x \right] 
	& \; = \; & \phantom{-}x
\nonumber \\
	\left[ \frac{\ri}{\hbar} \frac{xp + px}{2} , p \right] 
	& \; = \; & -p \; ,
\end{eqnarray}
so that the application of the transformation formula~(\ref{eq:BCH}) to 
the unitary operator 
\begin{equation}
	S_\lambda = 
	\exp\left(\frac{\ri}{\hbar} \ln\lambda \frac{xp+px}{2} \right)
\end{equation}
with arbitrary $\lambda > 0$ results in both	
\begin{equation}
	S_\lambda x \, S_\lambda^\dagger 
	\; = \; \sum_{n=0}^\infty \frac{(\ln\lambda)^n}{n!} x
	\; = \; \lambda x
\end{equation}	
and
\begin{equation}
	S_\lambda p \, S_\lambda^\dagger 
	\; = \; \sum_{n=0}^\infty \frac{(-\ln\lambda)^n}{n!} p
	\; = \; \frac{1}{\lambda} p \; .
\end{equation}	
Thus, $S_\lambda$ implements a scale transformation, leaving the product
$px$ invariant. If we now admit a time-dependent scaling parameter
$\lambda = \lambda(t)$, we also have 
\begin{equation}
	S_{\lambda(t)} 
	\left( -\ri\hbar \frac{\partial}{\partial t} \right) 
	S_{\lambda(t)}^\dagger =
	-\ri\hbar \frac{\partial}{\partial t} 
	-\frac{\dot{\lambda}(t)}{\lambda(t)} \, \frac{xp+px}{2} \; .
\end{equation}
Returning to Eq.~(\ref{eq:TR1}), this allows us to achieve two goals
simultaneously: Equating
\begin{equation}
	\frac{\dot{\lambda}(t)}{\lambda(t)} = 2\frac{\eta(t)}{M} \; ,
\end{equation}	
which by Eq.~(\ref{eq:FET}) is equivalent to
\begin{equation}
	\frac{\rd}{\rd t} \ln \lambda(t) = 
	\frac{\rd}{\rd t} \ln | \xi | \; ,
\label{eq:DLA}
\end{equation}
we may set 
\begin{equation}
	\lambda(t) = | \xi |
\label{eq:LAM}
\end{equation}
and define a second unitary transformation
\begin{equation}
	U_2 = S_{|\xi|} \; ,
\label{eq:UT2}
\end{equation}
effectuating
\begin{eqnarray}
	& & 
	U_2 U_1 \left( H_0(t) - \ri\hbar\frac{\partial}{\partial t} \right)
	U_1^\dagger U_2^\dagger
\nonumber \\	& = & 	
	U_2 \left( 
	\frac{p^2}{2M} + \frac{1}{2} M \Omega^2 \frac{x^2}{|\xi|^4}
	- \ri\hbar \frac{\partial}{\partial t} 
	+ \frac{\eta(t)}{M} \big( p x + x p \big) \right) U_2^\dagger
\nonumber \\ 	& = &
	\frac{1}{|\xi|^2} \left[ \frac{p^2}{2M} 
	+ \frac{1}{2}M\Omega^2 x^2 \right] 
	- \ri\hbar \frac{\partial}{\partial t} \; .
\label{eq:TR2}	
\end{eqnarray}	
Thus we have {\em both\/} scaled the momentum $p$ by $1/|\xi|$ and the 
position $x$ by $|\xi|$, allowing us to take a time-dependent factor 
$1/|\xi|^2$ out of the square brackets, {\em and\/} have removed the annoying
last term that had appeared on the right-hand side of Eq.~(\ref{eq:TR1}).	

Observe also that there is a further unexploited freedom: Integration 
of Eq.~(\ref{eq:DLA}) leaves us with an arbitrary constant $\ln c$, so that 
we might have chosen $\lambda(t) = c|\xi|$ instead of Eq.~(\ref{eq:LAM}).
This would have led to a renormalization of the mass in the last line of
Eq.~(\ref{eq:TR2}), shifting $M$ to $Mc^2$. As will become evident below, 
this freedom again has no observable consequences.

Now the result of the two-step transformation~(\ref{eq:TR2}) prompts us to 
solve the modified Schr\"odinger equation 
\begin{equation}
	\ri\hbar \frac{\partial}{\partial t} \chi(x,t)
	= \frac{1}{|\xi|^2} H_{\rm osc} \chi(x,t) \; ,
\label{eq:MSE}
\end{equation}
instead of Eq.~(\ref{eq:SGL}), where
\begin{equation}
	H_{\rm osc} = \frac{p^2}{2M} + \frac{1}{2}M\Omega^2 x^2
\label{eq:HOS}
\end{equation}
is the Hamiltonian of a {\em time-independent\/} harmonic 
oscillator~\cite{Husimi53,LewisRiesenfeld69,PopovPerelomov69}, possessing
the eigenfunctions
\begin{equation}
	\chi_n^{\rm osc}(x) = \frac{\pi^{-1/4}}{\sqrt{2^n n! \, L}} \,
	H_n(x/L) \, \exp\!\big(-(x/L)^2/2\big)
\label{eq:EIF}
\end{equation}	
with integer quantum numbers $n = 0, 1, 2, \ldots$, Hermite polynomials
$H_n$, and oscillator length $L = \sqrt{\hbar/(M\Omega)}$, yielding the
energy eigenvalues $E_n = \hbar\Omega(n + 1/2)$. Inserting the natural ansatz
\begin{equation}
	\chi_n(x,t) = \exp\!\big(-\ri\gamma_n(t)/\hbar\big) \, 
	\chi_n^{\rm osc}(x)
\label{eq:ANS}
\end{equation}
into Eq.~(\ref{eq:MSE}), one finds 
\begin{equation}
	\dot\gamma_n(t)  = \frac{1}{|\xi|^2} E_n \; .
\label{eq:GAM}
\end{equation}	
This equation for the desired phase $\gamma_n(t)$ can be brought into a 
more transparent form: Introducing the phase $\varphi(t)$ of the complex 
trajectory~$\xi$ according to
\begin{equation}
	\xi = | \xi | \exp\!\big(\ri \varphi(t) \big)
\end{equation}	
or
\begin{equation}
	\frac{\xi}{\xi^*} = \exp\!\big(2\ri \varphi(t) \big) \; ,
\end{equation}	
one derives
\begin{equation}
	\frac{\rd}{\rd t} \frac{\xi}{\xi^*} =
	\frac{\dot\xi \xi^* - \xi\dot\xi^*}{\xi^{*2}} =
	2\ri\dot\varphi(t) \, \exp\!\big(2\ri \varphi(t) \big) \; . 
\end{equation}
Again invoking the Wronskian~(\ref{eq:WRO}), this becomes
\begin{equation}
	\dot\varphi(t) = \frac{\Omega}{|\xi|^2} \; ;
\end{equation}		 
recall that this expression is not affected by the freedom to scale $\xi$
by an arbitrary factor. Hence, Eq.~(\ref{eq:GAM}) takes the form
\begin{equation}
	\dot\gamma_n(t) = \frac{E_n}{\Omega} \dot\varphi(t) \; ;
\end{equation}
observe that the frequency $\Omega$ of the auxiliary oscillator~(\ref{eq:HOS})
drops out here. Integrating, we have fully determined the 
solutions~(\ref{eq:ANS}) to Eq.~(\ref{eq:MSE}):
\begin{equation}
	\chi_n(x,t) = \exp\!\Big( -\ri(n + 1/2) 
	\big[ \varphi(t) - \varphi(0) \big] \Big) 
	\chi_n^{\rm osc}(x) \; ,
\label{eq:SOL}
\end{equation}	
having stipulated $\gamma(0) = 0$. The appearance of $\varphi(0)$ makes sure 
that these solutions~(\ref{eq:SOL}) remain invariant under a constant shift of 
the phase of $\xi$, as does the effective Hamiltonian $H_{\rm osc}/|\xi|^2$. 

Next, we need to invert the two transformations~(\ref{eq:UT2}) 
and~(\ref{eq:UT1}) in order to obtain the solutions of the original 
Schr\"odinger equation~(\ref{eq:SGL}). Utilizing the identity 
\begin{equation}
	S_\lambda f(x) = \sqrt{\lambda} f(\lambda x) \; ,
\end{equation}
which may be verified by differentiating both sides with respect to~$\lambda$, 
one finds
\begin{eqnarray}
\label{eq:GWF}
	\psi_n(x,t) & = & U_1^\dagger U_2^\dagger \, \chi_n(x,t)
\\	& = &
	\exp\!\Big( -\ri(n + 1/2) \big[ \varphi(t) - \varphi(0) \big] \Big)
	u_n^{\rm osc}(x,t) \; ,
\nonumber	
\end{eqnarray}
where
\begin{equation}	
	u_n^{\rm osc}(x,t) = \exp\!\left(\frac{\ri M}{2\hbar} x^2
	\frac{\rd}{\rd t} \ln |\xi| \right) 	
	\frac{1}{\sqrt{|\xi|}} 
	\chi_n^{\rm osc}\!\left(\frac{x}{|\xi|}\right) \; ;	
\label{eq:UOS}
\end{equation}		
of course, this expression agrees with the known solutions obtained by 
other approaches~\cite{Husimi53,LewisRiesenfeld69,PopovPerelomov69}. 
Had we utilized the freedom to choose $\lambda(t) = c|\xi|$ for the second 
transformation~(\ref{eq:UT2}), leading to the replacement of $M$ by $Mc^2$ 
in the last line of Eq.~(\ref{eq:TR2}), the oscillator length~$L$ would 
have been rescaled to $L/c$ in the eigenfunctions~(\ref{eq:EIF}), so that
the final results~(\ref{eq:GWF}) and~(\ref{eq:UOS}) remain unchanged. 

So far, these considerations apply to an {\em arbitrary} variation of the 
spring function with time. Now we require that $k$ depend {\em periodically\/}
on time with period~$T$,   
\begin{equation}
	k(t) = k(t+T) \; ,
\label{eq:TSC}	
\end{equation}	
so that the classical equation of motion~(\ref{eq:EOM}) becomes Hill's 
equation, which underlies the theory of parametric resonance, and therefore 
has been intensely studied~\cite{JoseSaletan98,MagnusWinkler04}. This equation 
possesses Floquet solutions, {\em i.e.\/}, solutions of the form 
\begin{equation}
	\xi(t) = v(t)\exp(\ri \nu t) \; , 
\label{eq:CFS}
\end{equation}
where the function $v(t)$ is periodic in time with the same period~$T$ as
the spring function,
\begin{equation}
	v(t) = v(t+T) \; .
\end{equation}		
The characteristic exponent $\nu$ can either be real, in which case $\xi$ 
and $\xi^*$ both constitute linearly independent {\em stable\/} solutions, 
or purely imaginary, in which case one of the two Floquet solutions grows
without bound and therefore is {\em unstable\/}, causing instability of the 
general solution~\cite{JoseSaletan98,MagnusWinkler04}. Here we restrict 
ourselves to the stable case, as this case allows one to construct 
normalized Floquet states of the parametrically driven quantum mechanical 
oscillator~\cite{PopovPerelomov70,Combescure86,Brown91}, that is, a 
complete set of solutions to the Schr\"odinger equation~(\ref{eq:SGL}) with 
time-periodic spring function~(\ref{eq:TSC}) having the particular guise      
\begin{equation}
	\psi_n(x,t) = u_n(x,t) \exp(-\ri\varepsilon_n t/\hbar) \; ,
\label{eq:FLO}
\end{equation}
where the Floquet functions
\begin{equation}
	u_n(x,t) = u_n(x,t+T)
\end{equation}	
again acquire the $T$-periodic time-dependence imposed by the spring function; 
the real quantities $\varepsilon_n$ are known as quasienergies. Indeed, 
inserting a stable classical Floquet solution~(\ref{eq:CFS}) into the wave 
functions~(\ref{eq:GWF}) obtained above, their factors $u_n^{\rm osc}(x,t)$ 
become $T$-periodic in time, since $|\xi| = |v|$. Moreover, writing
\begin{equation}
	v(t) = | v(t) | \exp\big(\ri \alpha(t) \big) \, ,   
\end{equation}
we necessarily have 
$\exp\!\big(\ri \alpha(t) \big) = \exp\!\big(\ri \alpha(t+T) \big)$,
and hence $\alpha(t+T) = \alpha(t) + 2\pi\ell$ with some integer winding
number~$\ell$. We then introduce
\begin{equation}
	\widetilde{\alpha}(t) = \alpha(t) - 2\pi\ell\frac{t}{T} \; ,
\label{eq:WIN}
\end{equation} 
implying that $\widetilde{\alpha}(t)$ actually is $T$-periodic,
$\widetilde{\alpha}(t) = \widetilde{\alpha}(t+T)$, and re-express 
Eq.~(\ref{eq:CFS}) in the form 
\begin{equation}
	\xi(t) = | v(t) | \, \re^{\ri \widetilde{\alpha}(t)} \,
	\exp\!\big(\ri(\nu + \ell\omega)t \big) \; ,
\label{eq:VAN}
\end{equation}	
where $\omega = 2\pi/T$. This representation~(\ref{eq:VAN}) brings out the 
content of the above steps more clearly: The factorization~(\ref{eq:CFS}) 
does not determine the characteristic exponent uniquely, but only up to an 
integer multiple of~$\omega$, 
\begin{equation}
	\nu \equiv \{ \nu + m\omega \; | \; 
	m \in {\mathbbm Z} \} \; . 
\label{eq:ECN}	
\end{equation}
Imposing the requirement that 
$v(t) = |v(t)| \exp\!\big( \ri \widetilde{\alpha}(t) \big)$ with $T$-periodic
phase function $\widetilde{\alpha}(t)$ then explicitly singles out one 
particular representative of this equivalence class~(\ref{eq:ECN}); this 
representative will be referred to as the {\em canonical representative\/} in 
the following. Adding the appropriate multiple of $\omega$ to the given $\nu$, 
we may henceforth adopt the convention that this canonical representative be 
labeled by $m = 0$.   

By the same token, instead of Eq.~(\ref{eq:FLO}) we could have written
\begin{equation}
	\psi_n(x,t) = u_n(x,t) \, \re^{\ri m\omega t}
	\exp\!\Big( - \ri(\varepsilon_n + m\hbar\omega)t/\hbar \Big)  
\end{equation}
with integer~$m$ and properly $T$-periodic Floquet functions 
$u_n(x,t) \exp(\ri m\omega t)$, signaling that a quasienergy likewise 
has to be regarded as a class of equivalent representatives,
\begin{equation}
	\varepsilon_n \equiv \{ \varepsilon_n + m\hbar\omega \; | \; 
	m \in {\mathbbm Z} \} \; . 	
\label{eq:QEC}
\end{equation}
After these preparations, the phase function $\varphi(t)$ appearing in the
solutions~(\ref{eq:GWF}) is identified as $\varphi(t) = \widetilde{\alpha}(t) 
+ \nu t$ in the case of a $T$-periodic spring function, with $\nu$ denoting 
the canonical representative of the characteristic exponent. Therefore, the 
$T$-periodic Floquet functions postulated by Eq.~(\ref{eq:FLO}) now coincide 
with the functions $u_n^{\rm osc}(x,t)$ up to a $T$-periodic phase factor,  
\begin{equation}
	u_n(x,t) = 
	\exp\!\Big( -\ri(n + 1/2) 
	\big[ \widetilde{\alpha}(t) - \widetilde{\alpha}(0) \big] \Big)
	u_n^{\rm osc}(x,t) \; ,	 
\label{eq:FFO}
\end{equation}
while their quasienergies are given by
\begin{equation}  
	\varepsilon_n = \hbar\nu(n + 1/2)
	\qquad \bmod \; \hbar\omega \; ,
\label{eq:QES}
\end{equation}
again writing $\omega = 2\pi/T$. Note that here the requirement that the 
phase function $\widetilde{\alpha}(t)$ of the periodic part $v(t)$ of the 
Floquet solutions~(\ref{eq:CFS}) itself be $T$-periodic selects particular, 
``canonical'' representatives of the quasienergy classes~(\ref{eq:QEC}). 
Thus, the quasienergy spectrum of the parametrically  driven harmonic 
oscillator~(\ref{eq:DHO}) does not depend on the parameters of the 
auxiliary oscillator~(\ref{eq:HOS}), which would be ill-defined anyway, 
but solely on the characteristic exponent~$\nu$ of the classical Floquet 
solution~(\ref{eq:CFS}), an observation which goes back to Popov and 
Perelomov~\cite{PopovPerelomov70}.

\section{Numerical example: The Mathieu oscillator}
\label{sec:3}

In order to construct Floquet solutions~(\ref{eq:CFS}) we re-write the 
classical equation of motion~(\ref{eq:EOM}) with $T$-periodic spring 
function~(\ref{eq:TSC}) as a system of two coupled first-order equations,
\begin{equation}
	\frac{\rd}{\rd t} 
	\left( \begin{array}{c} \xi \\ \dot\xi \end{array} \right) = 
	\left( \begin{array}{cc} 0 & 1 \\ -k(t)/M & 0 \end{array} \right)
	\left( \begin{array}{c} \xi \\ \dot\xi \end{array} \right) \; ,	
\end{equation}
and consider two solutions $\xi^{(1)}(t)$, $\xi^{(2)}(t)$ to this system
with the particular initial conditions
\begin{equation}
	\left( 
	\begin{array}{c} \xi^{(1)}(0) \\ \dot\xi^{(1)}(0) \end{array} 
	\right) = \left( \begin{array}{c} 1 \\ 0 \end{array} \right)
	\quad , \quad
	\left( 
	\begin{array}{c} \xi^{(2)}(0) \\ \dot\xi^{(2)}(0) \end{array} 
	\right) = \left( \begin{array}{c} 0 \\ 1 \end{array} \right) \; .
\end{equation}
By numerical integration we then obtain the one-cycle evolution matrix
\begin{equation}
	{\bm M} = \left( \begin{array}{cc}
	\xi^{(1)}(T) & \xi^{(2)}(T) \\
	\dot\xi^{(1)}(T) & \dot\xi^{(2)}(T) \end{array} \right) \; ,
\end{equation}
the eigenvalues of which constitute a pair of Floquet multi\-pliers 
$\exp(\pm\ri\vartheta)$~\cite{MagnusWinkler04}. In the stable case, that is,
when $\vartheta$ turns out to be real, these Floquet multipliers both lie on 
the unit circle, giving the characteristic exponent
\begin{equation}
	\nu = \pm\vartheta/T \qquad \bmod \; \omega \; ,
\end{equation}
leaving the selection of the canonical representative still open. Moreover, 
with $(y_1, y_2)^{\rm t}$ denoting an eigenvector of ${\bm M}$ belonging 
to one of the eigenvalues $\exp(\pm\ri\vartheta)$, the required Floquet 
solutions~(\ref{eq:CFS}) are given by
\begin{equation}
	\xi(t) = y_1 \, \xi^{(1)}(t) + y_2 \, \xi^{(2)}(t) \; . 
\end{equation}

We now apply this machinery to the particular function 
\begin{equation}
	k(t) = M \Omega_0^2 - M \Omega_1^2 \cos(\omega t) \; .	
\label{eq:MSF}
\end{equation}
Invoking the dimensionless time variable $\omega t = 2\tau$, Hill's
equation~(\ref{eq:EOM}) then becomes equal to the Mathieu equation in its 
standard form~\cite{AbramowitzStegun70},
\begin{equation}
	\frac{\rd^2}{\rd\tau^2}\xi + \big[ a - 2 q \cos(2\tau) \big] \xi = 0
	\; ,
\label{eq:MAE}
\end{equation}
with parameters
\begin{eqnarray}
	a & = & \frac{4\Omega_0^2}{\omega^2}
\nonumber \\
	q & = & \frac{2\Omega_1^2}{\omega^2} \; .
\label{eq:PAR}
\end{eqnarray}	 	
Among others, this Mathieu equation~(\ref{eq:MAE}) underlies the conception 
of mass spectrometers without magnetic fields~\cite{PaulSteinwedel53}, and 
the design of the Paul trap~\cite{Dehmelt90,Paul90}. Thus, the parametrically 
driven harmonic oscillator with the spring function~(\ref{eq:MSF}) will be 
referred to as the {\em Mathieu oscillator\/}. 

This particular example now allows us to substantiate the choice of the 
canonical representative of the characteristic exponent. Namely, when the 
scaled driving strength~$q$ defined by Eq.~(\ref{eq:PAR}) goes to zero, Hill's 
equation reduces to the equation of motion for a classical harmonic oscillator 
with frequency~$\Omega_0$, providing solutions $\xi(t) = \exp(\pm\ri\Omega_0 t)$
and, hence, $\nu(q = 0) = \Omega_0 \bmod \omega$. In order to make sure that 
the canonical representatives of the quasienergies~(\ref{eq:QES}) actually 
connect to the quantum mechanical energy eigenvalues $E_n = \hbar\Omega_0
(n + 1/2)$ of such an undriven oscillator, we impose the condition that 
$\nu \to \Omega_0$ in this limit $q \to 0$. Starting from the eigenvalues 
$\exp(\pm\ri\vartheta) = \exp(\pm\ri\Omega_0 T)$ of ${\bm M}(q=0)$, with
$0 < \vartheta < \pi$, and denoting the integer part of the ratio 
$\Omega_0/\omega$ by ${\rm int}(\Omega_0/\omega) = \ell_0$, we have 
\begin{equation}
	\frac{\nu(q=0)}{\omega} = \left\{ \begin{array}{lll}
		\ell_0 + \difrac{\vartheta}{2\pi} & \;{\rm if} \; & 
		0 < \difrac{\Omega_0}{\omega} - \ell_0 < \difrac{1}{2} \; , \\
		\ell_0 + 1 - \difrac{\vartheta}{2\pi} & \;{\rm if} \; &
		\difrac{1}{2} < \difrac{\Omega_0}{\omega} - \ell_0 < 1 \; . 
	\end{array} \right.	
\label{eq:AQZ}
\end{equation}
Thus, the function $\widetilde\alpha(t)$ appearing in Eq.~(\ref{eq:WIN}) 
is identically equal to zero for $q = 0$. This assignment~(\ref{eq:AQZ}), 
unambiguously made at $q = 0$, is then extended to the entire zone of 
stability connected to the parameters $a = 4\Omega_0^2/\omega^2$ and $q = 0$ 
by continuity.

\begin{figure}[t]
\centering
\includegraphics[width=0.9\linewidth]{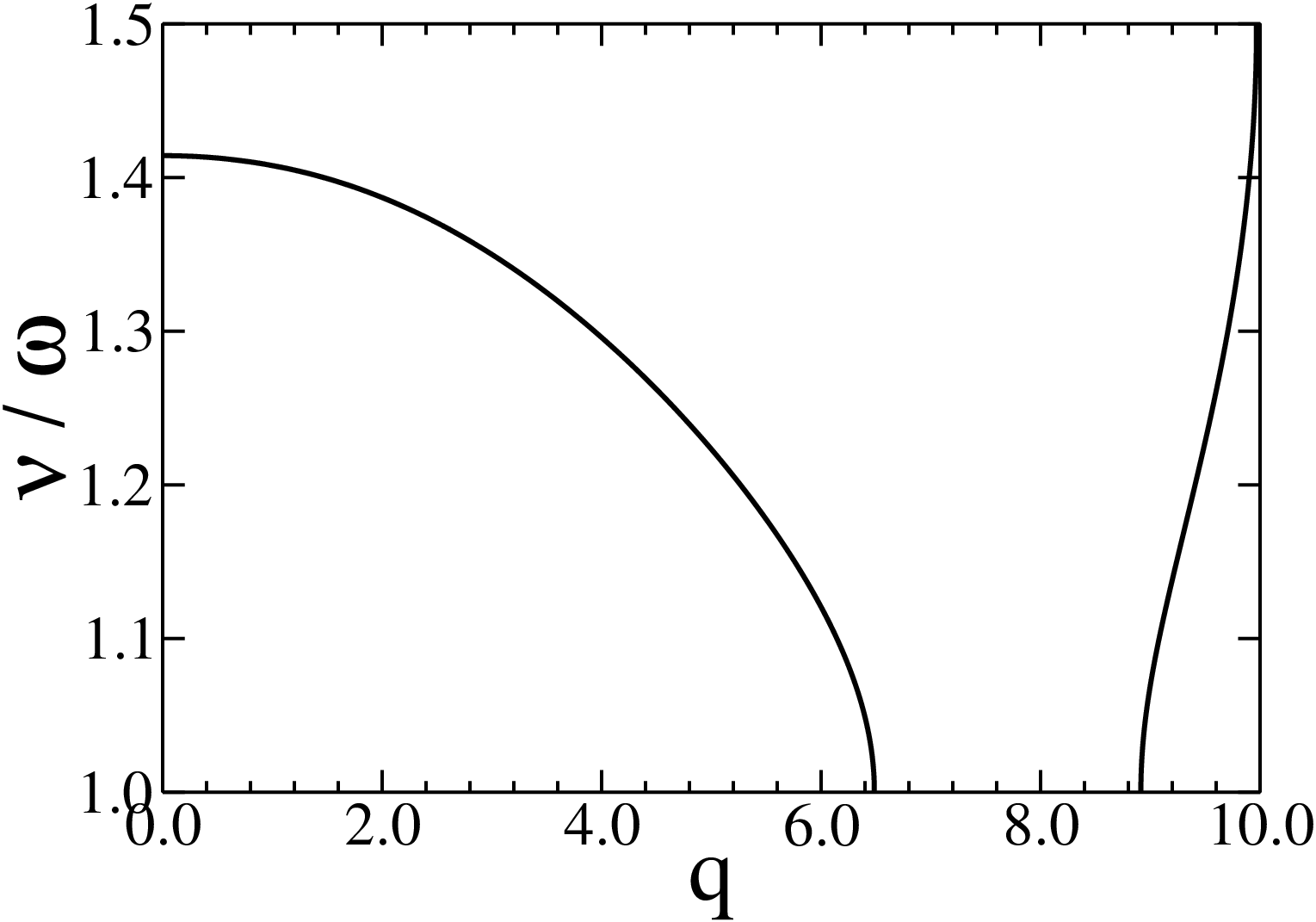}
\caption{Canonical representative of the characteristic exponent~$\nu$ for
	the Mathieu oscillator with parameter $a = 8.0$, that is, for
	$\Omega_0/\omega = \sqrt{2}$, as function of the scaled driving
	strength $q = 2\Omega_1^2/\omega^2$. The oscillator becomes unstable
	for $q \approx 6.49$, where $\nu/\omega = 1$; then re-enters a regime 
	of stability at $q \approx 8.91$, and becomes unstable again at 
	$q \approx 9.97$, where $\nu/\omega = 3/2$.}
\label{F_1}
\end{figure}

\begin{figure}[th]
\centering
\includegraphics[width=0.9\linewidth]{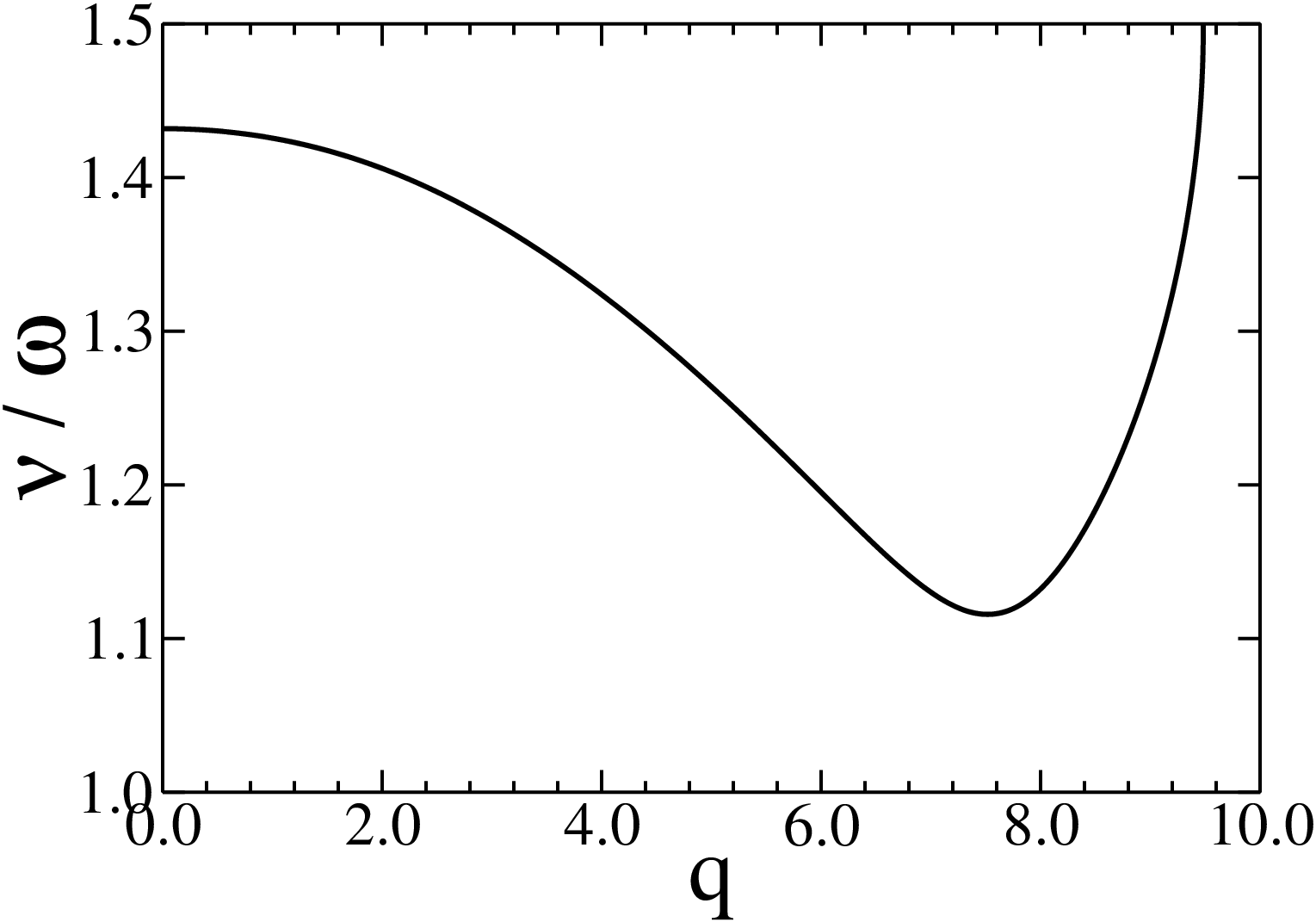}
\caption{As Fig.~\ref{F_1}, but for $a = 4\Omega_0^2/\omega^2 = 8.2$. 
	Here the oscillator becomes unstable for $q \approx 9.48$, where 
	$\nu/\omega = 3/2$.}
\label{F_2}
\end{figure}

The Figures~\ref{F_1} and~\ref{F_2} visualize the variation of the 
characteristic exponent with the scaled driving strength~$q$ for $a = 8.0$ and 
$a = 8.2$, respectively; in view of Eq.~(\ref{eq:QES}) for the quasienergies,
these Figures likewise depict the ac Stark shift exhibited by the quantum
mechanical Mathieu oscillator.

It is of interest to observe that the classical Mathieu oscillator becomes 
mechanically unstable upon variation of~$q$ in two different ways: Two complex 
eigenvalues $z_{\pm} = \exp(\pm\ri\vartheta)$ of ${\bm M}$ collide on the unit 
circle and become real either when $z_{\pm}= +1$, so that $\vartheta = 0$ and 
$\nu/\omega$ is integer, or when $z_{\pm}= -1$, giving $\vartheta = \pi$ and 
half-integer $\nu/\omega$. In the first case all quasienergies~(\ref{eq:QES}) 
of the corresponding quantum system are degenerate ($\bmod\;\hbar\omega$) at 
the transition point, whereas there are two separate groups of degenerate 
quasienergies, differing by $\hbar\omega/2$, in the second case, thus providing
an elementary model for a Floquet time crystal~\cite{ElseEtAl16,YaoEtAl17}.
In fact, the transition from a stable to an unstable classical Mathieu 
oscillator corresponds to a transition from a pure point quasienergy 
spectrum to an absolutely continuous one for its quantum mechanical 
counterpart~\cite{HagedornEtAl86,Combescure88,Howland92}. In a regime of 
stability each wave function possesses a representation as a superposition 
of the Floquet states constructed in Sec.~\ref{sec:2}, and therefore evolves 
in time in a strictly quasiperiodic manner. In a regime of instability the
solutions to the time-dependent Schr\"odinger equation are still associated 
with classical trajectories~$\xi$, but these wave functions absorb an infinite 
amount of energy from the drive if their trajectories are unstable; such 
unbounded growth of energy is traced to the continuous quasienergy 
spectrum~\cite{BunimovichEtAl91}.

\section{Coupling to a thermal heat bath}
\label{sec:4}

Now let the parametrically driven ``system''~(\ref{eq:DHO}) with $T$-periodic 
spring function~$k(t)$ be weakly coupled to a ``bath'' consisting of infinitely
many harmonic oscillators with a prescribed temperature, causing transitions 
among the system's Floquet states; our goal is to find the corresponding 
quasistationary distribution~\cite{BreuerEtAl00,LangemeyerHolthaus14,
SchmidtEtAl19,KetzmerickWustmann10}. Following the general theory of open 
quantum systems~\cite{BreuerPetruccione02} we then require, besides the Hilbert
space ${\mathcal H}_{\rm system}$ that the driven part $H_0(t)$ is acting on, 
the Hilbert space ${\mathcal H}_{\rm bath}$ pertaining to the bath Hamiltonian 
$H_{\rm bath}$, and construct the composite space 
${\mathcal H}_{\rm system} \otimes {\mathcal H}_{\rm bath}$. Accordingly,
the  total Hamiltonian now takes the form
\begin{equation}
	H(t) = H_0(t) \otimes \mathbbm{1} + \mathbbm{1} \otimes H_{\rm bath}
	+ H_{\rm int} \; , 
\end{equation}
with
\begin{equation}
	H_{\rm int} = V \otimes W
\end{equation}	
specifying the system-bath interaction. Here we choose a simple but
plausible coupling mediated by   
\begin{equation}
	V = \gamma x \; ,
\label{eq:DTC}
\end{equation}	
where the constant $\gamma$ carries the dimension of energy per length, 
and
\begin{equation}
	W = \sum_{\widetilde\omega} 
	\left( b^{\phantom\dagger}_{\widetilde\omega}
	+ b^\dagger_{\widetilde\omega} \right) \; ,
\end{equation}	
effectuating annihilation and creation processes in the bath, with the sum
ranging over all bath oscillators~\cite{BreuerEtAl00}. Within a perturbative
approach based on the golden rule for Floquet states~\cite{BreuerEtAl00,
LangemeyerHolthaus14}, a bath-induced transition from an initial Floquet 
state~$i$ of the driven system to a final one labeled by~$f$ does not 
correspond to only one single transition frequency, but rather to an infinite 
ladder of frequencies $\omega_{fi}^{(\ell)}$ differing from the expected 
frequency $(\varepsilon_f - \varepsilon_i)/\hbar$ by positive or negative 
integer multiples of the driving frequency $\omega = 2\pi/T$,
\begin{equation}
	\omega_{fi}^{(\ell)} = 
	(\varepsilon_f - \varepsilon_i)/\hbar + \ell\omega 
	\quad \text{with} \; \ell = 0,\pm 1, \pm 2, \ldots \; ;
\label{eq:TRF}
\end{equation}
note that a precise specification of the chosen quasienergy representatives 
is essential at this point. Hence, the rate $\Gamma_{fi}$ of such transitions
is obtained as a sum,
\begin{equation}
	\Gamma_{fi} = \sum_\ell \Gamma_{fi}^{(\ell)} \; ,
\label{eq:GFI}
\end{equation}
where the partial rates~$\Gamma_{fi}^{(\ell)}$ are given by
\begin{equation}
	\Gamma_{fi}^{(\ell)} = 
	\frac{2\pi}{\hbar^2} \left| V_{fi}^{(\ell)} \right|^2 
	N(\omega_{fi}^{(\ell)}) \, J(|\omega_{fi}^{(\ell)}|) \; .
\label{eq:PRA}
\end{equation}
Here the quantities $V_{fi}^{(\ell)}$ denote the Fourier coefficients of 
the system's transition matrix elements,
\begin{equation}
	\langle u_f(t) | V | u_i(t) \rangle =
	\sum_{\ell } \re^{\ri \ell \omega t} \, V_{fi}^{(\ell)} \; .
\label{eq:TME}
\end{equation}
The transition frequencies~(\ref{eq:TRF}) can either be positive, as 
corresponding to processes during which the driven system absorbs energy 
from the bath, or negative, so that the system loses energy to the bath. 
Accordingly, the thermal averages $N(\widetilde\omega)$ appearing in the 
partial rates~(\ref{eq:PRA}) either refer to the de-excitation of a bath 
oscillator, that is, to the annihilation of a bath phonon,
\begin{equation}
	N(\widetilde\omega) \; = \; \langle n(\widetilde\omega) \rangle
	\; = \; \frac{1}{\exp(\beta\hbar\widetilde\omega) - 1}
\label{eq:BOP}
\end{equation}
when $\widetilde\omega > 0$, or to the creation of such a phonon,
\begin{equation}
	N(\widetilde\omega) \; = \; \langle n(-\widetilde\omega) \rangle + 1 
	\; = \; \frac{1}{1 - \exp(\beta\hbar\widetilde\omega)}
\label{eq:BOM}
\end{equation}
when $\widetilde\omega < 0$. Here we have written $n(\widetilde\omega) = 
b^\dagger_{\widetilde\omega} b^{\phantom\dagger}_{\widetilde\omega}$ 
for the occupation number of a phonon mode, have employed angular brackets 
to indicate thermal averaging, and have used the familiar symbol 
$\beta = 1/(k_{\rm B}T_{\rm bath})$ with the Boltzmann constant~$k_{\rm B}$ 
to indicate the inverse of the bath temperature $T_{\rm bath}$. Finally, the 
factor $J(\widetilde\omega)$ contributing to the partial rates~(\ref{eq:PRA}) 
denotes the spectral density of the oscillator bath.

Having computed the matrix of transition rates~(\ref{eq:GFI}) in this manner, 
the desired quasistationary distribution $\{ p_n \}_{n = 0,1,2,\ldots}$ 
which quantifies the system's Floquet-state occupation probabilities in the 
non-equilibrium steady state is obtained as solution to the Pauli master 
equation~\cite{BreuerEtAl00}   
\begin{equation}
	\dot{p}_n = 0 = \sum_m \big( \Gamma_{nm}p_m - \Gamma_{mn} p_n \big)
	\; .
\label{eq:PME}
\end{equation}	 			 					

The decisive system-specific input data determining this quasistationary 
Floquet-state distribution are the Fourier coefficients of the transition 
matrix 
elements~(\ref{eq:TME}). With the dipole-type coupling~(\ref{eq:DTC}), and 
again writing the decomposition of the $T$-periodic factor~$v(t)$ of the
classical Floquet solutions~(\ref{eq:CFS}) as
\begin{equation}
	v(t) = | v(t) | \exp\big(\ri \widetilde\alpha(t) \big)	
\end{equation}
with $T$-periodic phase function $\widetilde\alpha(t)$, the quantum mechanical 
Floquet functions~(\ref{eq:FFO}) of the parametrically driven harmonic 
oscillator provide the expression 
\begin{widetext}
\begin{eqnarray}
\label{eq:DME}
	\langle u_m(t) | x | u_n(t) \rangle & = & | v(t)| \left\langle u_m(t) 
	\left| \frac{x}{|v(t)|} \right| u_n(t) \right\rangle
\nonumber \\	& = &
	\sqrt{\frac{\hbar}{2M\Omega}} \, |v(t)| 
	\Big( \sqrt{n} \exp\!\big(-\ri\widetilde\alpha(t)
	+ \ri\widetilde\alpha(0) \big) \, \delta_{m,n-1}	 
	+ \sqrt{n+1} \exp\!\big(\ri\widetilde\alpha(t)
	- \ri\widetilde\alpha(0) \big) \, \delta_{m,n+1} \Big)
\nonumber \\	& = &
	\sqrt{\frac{\hbar}{2M\Omega}}
	\Big( \sqrt{n} \, 
	v^*(t) \, \re^{\ri\widetilde\alpha(0)}\, \delta_{m,n-1} 
	+ \sqrt{n+1} \, v(t) \, \re^{-\ri\widetilde\alpha(0)} \, 
	\delta_{m,n+1} \Big) \; .				
\end{eqnarray}
\end{widetext}
Therefore, the required coefficients~$V_{fi}^{(\ell)}$ of the 
expansion~(\ref{eq:TME}) are proportional to the Fourier coefficients 
of $v(t)$, which are easy to compute. Moreover, the transition 
matrix~(\ref{eq:GFI}) becomes tridiagonal, having non-vanishing entries for 
$f = i \pm 1$ only. Thus, the master equation~(\ref{eq:PME}) simplifies
considerably, reducing to
 \begin{eqnarray}
	( \Gamma_{n,n-1} \, p_{n-1} - \Gamma_{n-1,n} \, p_n ) & &
\nonumber \\	
      + \; ( \Gamma_{n,n+1} \, p_{n+1} - \Gamma_{n+1,n} \, p_n ) & = & 0
\label{eq:TRI}
\end{eqnarray} 
for $n \ge 1$; for $n = 0$ the first bracket disappears. This tridiagonal
form implies detailed balance~\cite{SchmidtEtAl19}, meaning that both brackets 
vanish individually for $n \ge 1$: Setting the second bracket to zero gives the
forward relation
\begin{equation}
	\frac{p_{n+1}}{p_n} = \frac{\Gamma_{n+1,n}}{\Gamma_{n,n+1}} 
\label{eq:FRE}
\end{equation}
which already fixes the distribution $\{ p_n \}_{n = 0,1,2,\ldots}$ up to 
its normalization; shifting $n$ to $n-1$ in this relation~(\ref{eq:FRE}) 
shows that Eq.~(\ref{eq:TRI}) indeed is satisfied. Moreover, since 
$\Gamma_{n+1,n}$ and $\Gamma_{n,n+1}$ both are proportional to $n+1$ 
by virtue of Eq.~(\ref{eq:DME}), their ratio
\begin{equation}
	\frac{\Gamma_{n+1,n}}{\Gamma_{n,n+1}} = r
\end{equation}
actually is independent of $n$, resulting in the geometric Floquet-state 
distribution
\begin{equation}
	p_n = (1 - r) \, r^n \; 
\label{eq:GEO}
\end{equation}
with the proviso that $r < 1$, that is, provided the rate $\Gamma_{n+1,n}$ 
for each ``upward'' transition remains smaller than the rate $\Gamma_{n,n+1}$ 
of the matching ``downward'' transition. As in the case of a spin driven by 
a circularly polarized field~\cite{SchmidtEtAl19}, the existence of such a 
geometric distribution~(\ref{eq:GEO}), combined with equidistantly spaced
canonical representatives~(\ref{eq:QES}) of the system's quasienergies, now 
allows one to introduce a quasitemperature~$\tau$ for the periodically driven
nonequilibrium system: Setting
\begin{equation}
	r = \exp\!\left(-\frac{\hbar\nu}{\kB\tau}\right) \; ,  
\end{equation}
one finds
\begin{equation}
	\frac{\tau}{T_{\rm bath}} = -\frac{\hbar\nu}{\kB T_{\rm bath}} \, 
	\frac{1}{\ln r} \; .		 
\end{equation}
This definition of the quasitemperature formally yields negative $\tau$
when $r > 1$. While such negative quasi\-temperatures are quite natural and 
physically meaningful in systems with a finite-dimensional Hilbert space, such 
as periodically driven spin systems~\cite{SchmidtEtAl19}, here they signal 
{\em quasithermal instability\/}, implying $\Gamma_{n+1,n} > \Gamma_{n,n+1}$, 
so that the particle tends to climb the oscillator ladder to infinite height.  

Writing the Fourier series of $v(t)$ as
\begin{equation}
	v(t) = \sum_\ell \re^{\ri\ell\omega t} \, v^{(\ell)} \; ,   
\label{eq:FSV}
\end{equation}
one has, more explicitly, 
\begin{equation}
	r = \frac{\displaystyle{\sum_\ell} \left| v^{(\ell)} \right|^2 \, 
		N(+\nu + \ell\omega) \; J(|\nu + \ell\omega |)}
	         {\displaystyle{\sum_\ell} \left| v^{(\ell)} \right|^2 \, 
		N(-\nu - \ell\omega) \; J(|\nu + \ell\omega |)} \; . 
\label{eq:EXR}
\end{equation}
Let us now assume that the system approaches a mechanical stability border, 
such that $\nu/\omega$ tends to the integer~$l_0$ from above,
$\nu/\omega \to \ell_0$. In that case the Bose occupation numbers 
$N(+\nu -\ell_0\omega) = N(0+)$ and $N(-\nu + \ell_0\omega) = N(0-)$ both 
become singular according to their respective definition~(\ref{eq:BOP}) or 
(\ref{eq:BOM}), with $N(0-) = N(0+) + 1$. Hence, assuming further that the
Fourier coefficient labeled $-\ell_0$ is of appreciable magnitude, and the
spectral density $J(|\widetilde\omega|)$ smoothly approaches a nonvanishing 
value $J(0)$, both the numerator and the denominator of the 
expression~(\ref{eq:EXR}) are practically exhausted by the term 
$\ell = -\ell_0$ alone, resulting in 
\begin{equation}
	r \approx \frac{N(0+)}{N(0+) + 1}
\end{equation}
and, hence, $r \to 1$, implying $\tau \to \infty$: If the 
(smooth) spectral density of the oscillator bath does not vanish at 
$\widetilde\omega = 0$, the onset of mechanical instability for integer 
$\nu/\omega$ necessarily is accompanied by quasi\-thermal instability.

Actually this link between mechanical and quasi\-thermal instability is even 
closer: If the spring function $k(t)$ admits symmetric or antisymmetric
functions $v(t)$ at the mechanical stability border, as it happens in the
Mathieu case~(\ref{eq:MSF}), one finds
\begin{equation}
	\left| v^{(+\ell-\ell_0)} \right|^2 =
	\left| v^{(-\ell-\ell_0)} \right|^2
	\quad {\rm if} \; \nu/\omega \to \ell_0 \; ,
\end{equation}
or 
\begin{equation}
	\left| v^{(+\ell-\ell_0)} \right|^2 = 
	\left| v^{(-\ell-\ell_0-1)} \right|^2 	
	\quad {\rm if} \; \nu/\omega \to \ell_0 + 1/2 \; .
\end{equation}
In both limiting cases the sums in the numerator and denominator of the
ratio~(\ref{eq:EXR}) become identical, so that the onset of mechanical 
instability entails $\tau = \infty$, even {\em regardless\/} of the bath 
density.

\begin{figure}[t]
\centering
\includegraphics[width=0.9\linewidth]{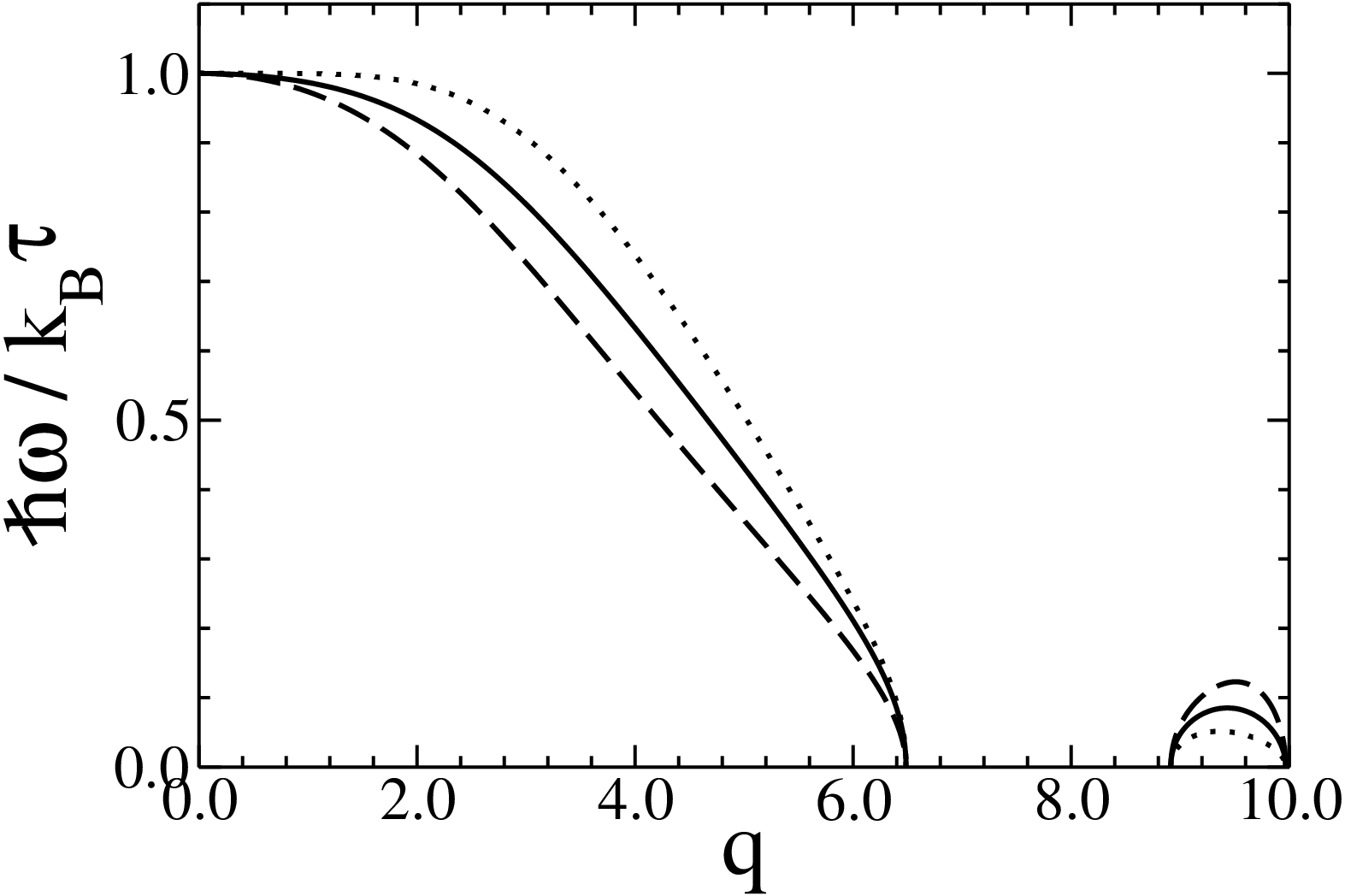}

\vspace{4.5ex}

\includegraphics[width=0.9\linewidth]{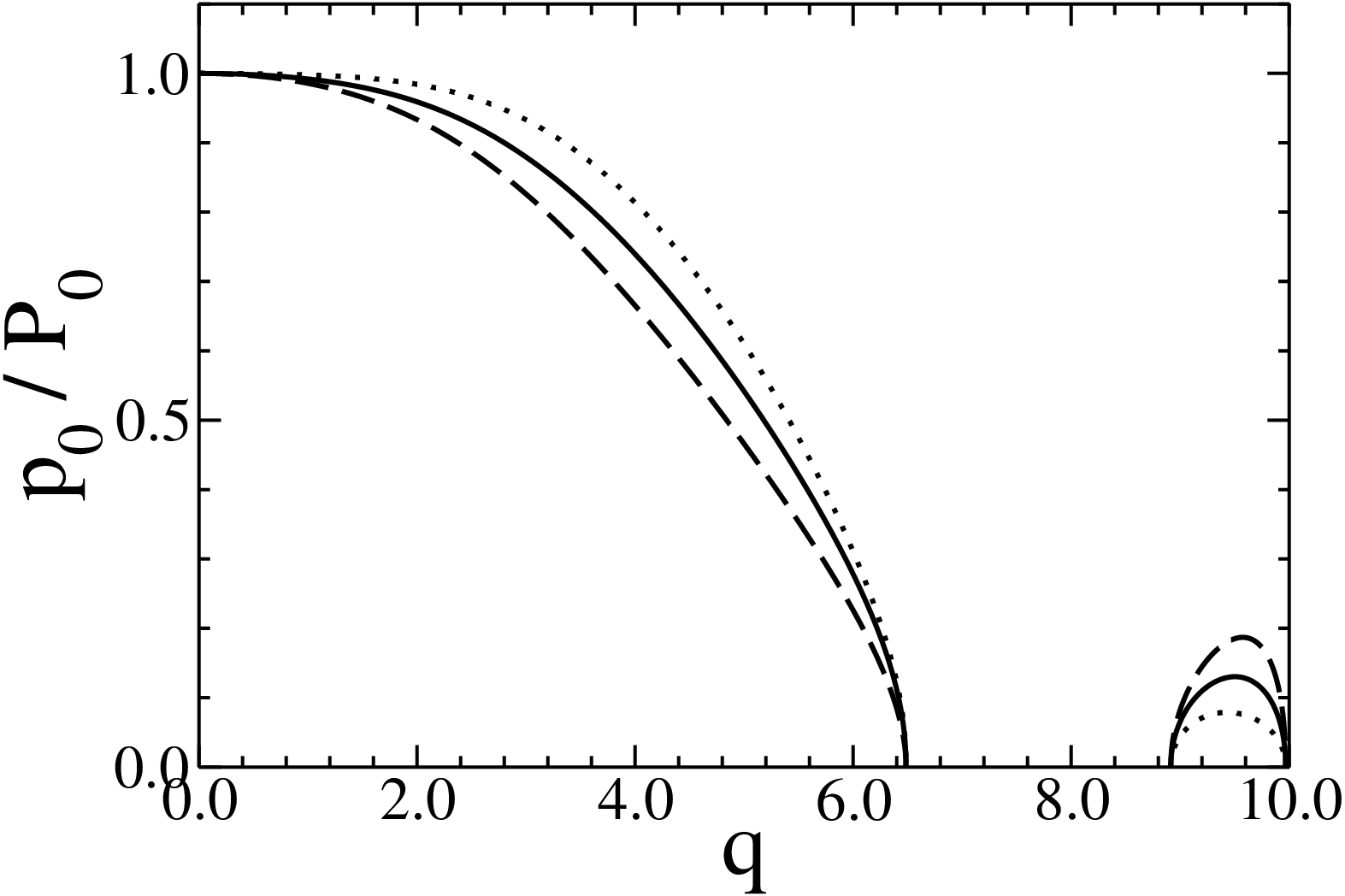}
\caption{Scaled inverse quasitemperature $\hbar\omega/(\kB\tau)$ (upper panel) 
	and ratio $p_0/P_0$ (lower panel) of the occupation probability~$p_0$ 
	of the Floquet state $n = 0$ to the occupation probability~$P_0$ of 
	the undriven system's ground state for a Mathieu oscillator with 
	$a = 8.0$, as functions of the scaled driving strength~$q$. The bath 
	temperature corresponds to $\beta\hbar\omega = 1$, while the spectral
	density of the bath is Ohmic ($s = 1$, full lines), sub-Ohmic 
	($s = 0.5$, dashed lines), and super-Ohmic ($s = 2$, dotted lines). 
	Observe that one has infinite quasitemperature at each mechanical 
	stability border seen in Fig.~\ref{F_1}.}  
\label{F_3}
\end{figure}

For illustrating these deliberations we resort once again to the Mathieu
oscillator~(\ref{eq:MSF}), and now stipulate that the spectral density
has the power-law form
\begin{equation}
	J(\widetilde\omega) = J_0 
	\left(\frac{\widetilde\omega_{\phantom 0}}
	           {\widetilde\omega_0}\right)^s \; .
\label{eq:OHM}
\end{equation}
The case $s = 1$ is designated as Ohmic~\cite{BreuerPetruccione02}, so that
exponents $0< s < 1$ and $s > 1$ indicate, respectively, sub-Ohmic and
super-Ohmic densities. In Fig.~\ref{F_3} we display the scaled inverse 
quasitemperature $\hbar\omega/(\kB\tau)$ as function of the scaled driving 
strength~$q$ for all three cases, considering an oscillator with $a = 8.0$ 
as in Fig.~\ref{F_1}, while the bath temperature has been set to 
$\beta\hbar\omega = 1.0$. We also plot the ratio $p_0/P_0$ of the quasithermal 
occupation probability 
\begin{equation}
	p_0 = 1 - r
\end{equation}	 
of the Floquet state $n = 0$ to the thermal occupation probability of the 
undriven oscillator's ground state,
\begin{equation}  
	P_0 = 1 - \exp(-\beta\hbar\Omega_0) \; .
\end{equation}	
Here the scaled inverse quasitemperature falls below the inverse bath 
temperature in both stable regions, implying that the driven system with 
$q > 0$ effectively is {\em hotter\/} than the undriven one with $q = 0$, so 
that the occupation probability of the Floquet state $n = 0$ is lower than 
the occupation probability of the oscillator ground state in the absence of 
the drive, as might be expected naively on intuitive grounds. Moreover, each 
border of mechanical stability identified before in Fig.~\ref{F_1} precisely 
marks an onset of quasithermal instability, {\em i.e.\/}, a driving strength 
for which $r = 1$, or $\hbar\omega/(\kB\tau) = 0$.

\begin{figure}[t]
\centering
\includegraphics[width=0.9\linewidth]{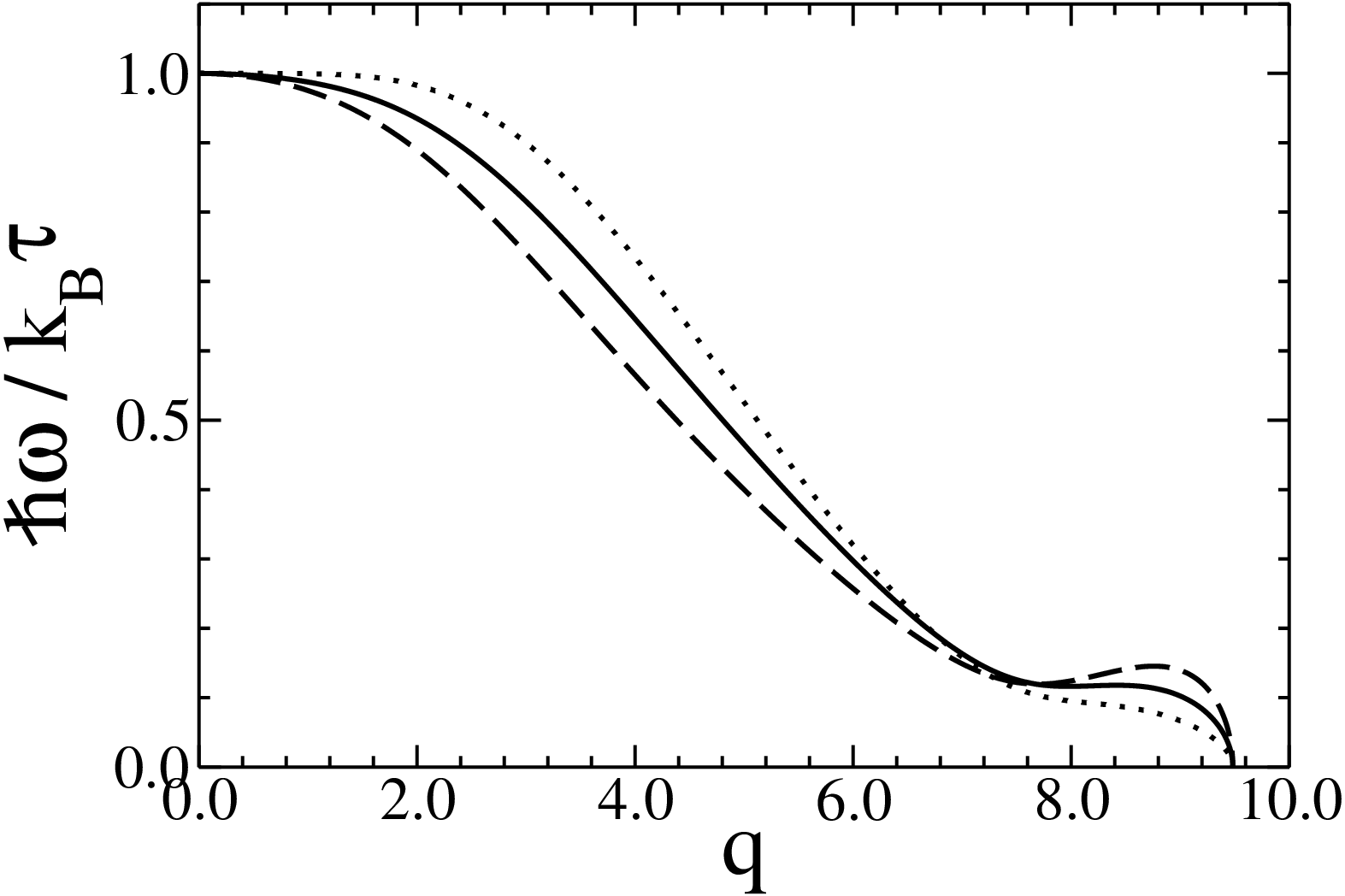}

\vspace{4.5ex}

\includegraphics[width=0.9\linewidth]{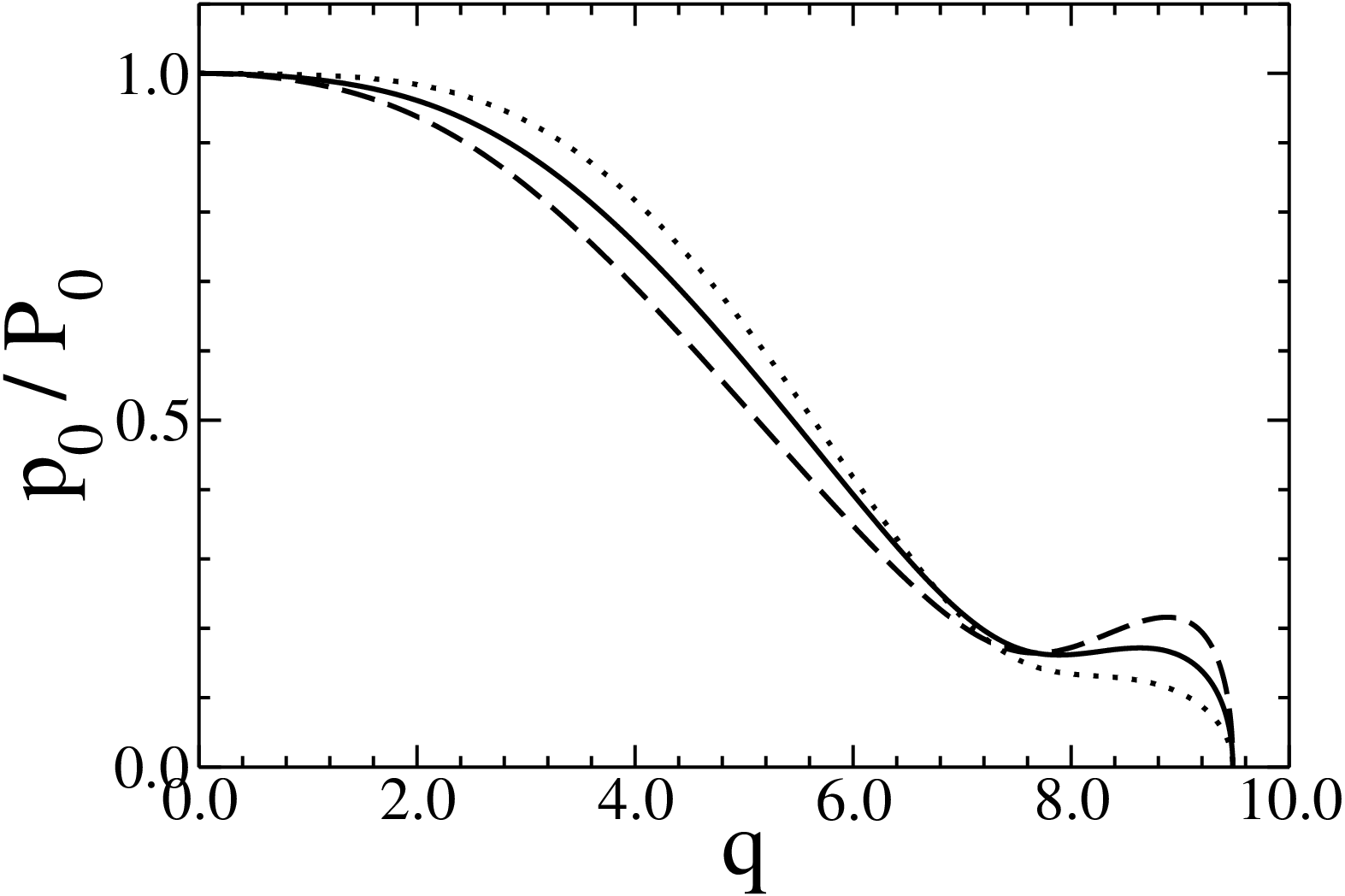}
\caption{As Fig.~\ref{F_3}, but for $a = 8.2$. Observe that the sub-Ohmic
	density of states gives rise to a non-monotonic variation of the 
	quasitemperature with the driving strength. } 
\label{F_4}
\end{figure}

The corresponding data for $a = 8.2$ are shown in Fig.~\ref{F_4}. Here the 
sub-Ohmic density gives rise to a regime in which the inverse quasitemperature 
{\em increases\/} notably with increasing driving strength, similar to the 
second zone of stability in Fig.~\ref{F_3}, indicating that the system can 
effectively become {\em colder\/} though the drive is made {\em stronger\/}, 
reflecting the behavior of the characteristic exponent depicted in 
Fig.~\ref{F_2}.

\begin{figure}[t]
\centering
\includegraphics[width=0.9\linewidth]{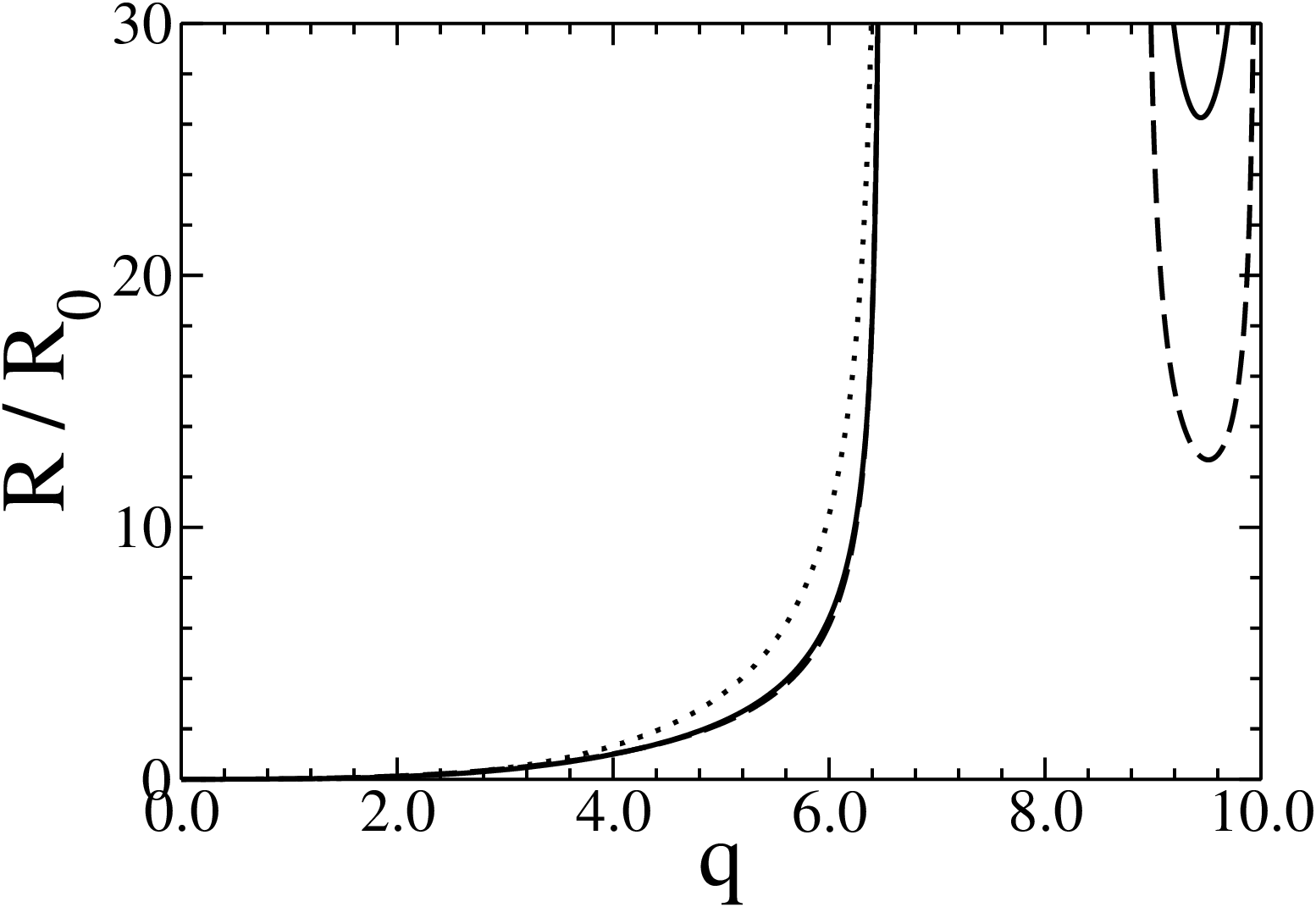}

\vspace{4.5ex}

\includegraphics[width=0.9\linewidth]{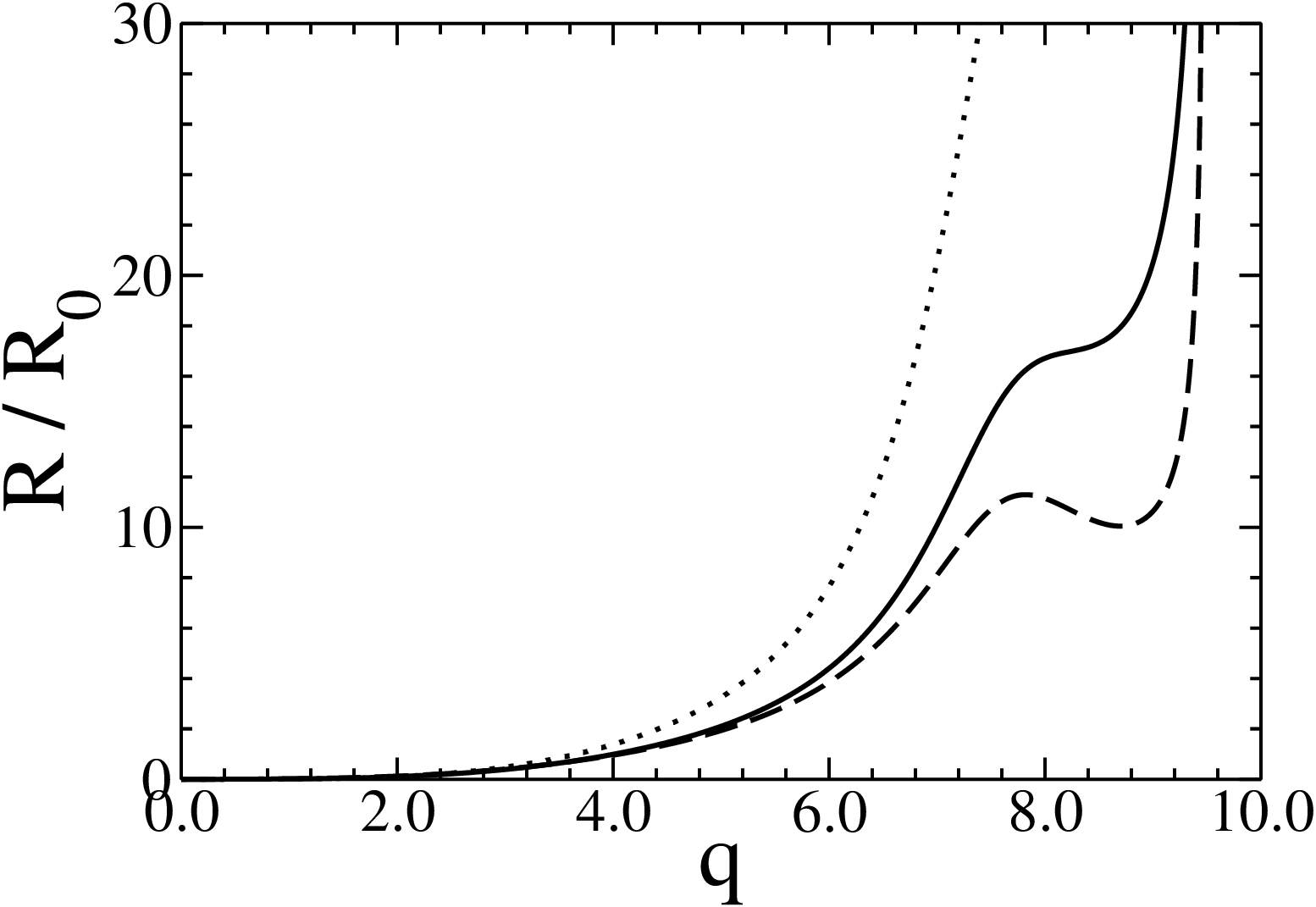}
\caption{Scaled dissipation rates $R/R_0$ for a Mathieu oscillator with
	$a = 8.0$ (upper panel) and $a = 8.2$ (lower panel) as functions of
	the driving strength. As in Figs.~\ref{F_3} and~\ref{F_4}, the bath 
	temperature  corresponds to $\beta\hbar\omega = 1$, while the spectral
	density of the bath is Ohmic ($s = 1$, full lines), sub-Ohmic 
	($s = 0.5$, dashed lines), and super-Ohmic ($s = 2$, dotted lines).
	The reference frequency entering the spectral densities~(\ref{eq:OHM})
	here is $\widetilde\omega_0/\omega = 1.0$.}  
\label{F_5}
\end{figure}

Since a Floquet transition of the system with negative (or positive) 
frequency $\omega_{mn}^{(\ell)}$ is accompanied by the addition (or 
subtraction) of the energy $-\hbar\omega_{mn}^{(\ell)}$ to (or from) 
the bath, the rate of energy dissipated in the quasi\-stationary state 
is given by~\cite{LangemeyerHolthaus14} 
\begin{equation}
	R = -\sum_{mn\ell} 
	\hbar\omega_{mn}^{(\ell)} \, \Gamma_{mn}^{(\ell)} \; p_n \; .
\label{eq:DIS}
\end{equation}	
Utilizing $m = n \pm 1$ together with $\omega_{n\pm1,n}^{(\ell)} = 
\pm \nu + \ell\omega$, and introducing the constant 
\begin{equation}
	c = \frac{\pi\gamma^2}{\hbar M\Omega}
\end{equation}	
which carries the dimension of squared inverse time, this dissipation 
rate~(\ref{eq:DIS}) can be written as 
\begin{equation}
	R = R_1 + R_2 \; ,
\label{eq:R12}
\end{equation}	
where 
\begin{equation}
	R_1 = \frac{cr}{1-r} \sum_\ell
	\hbar |\nu + \ell\omega| \, \left| v^{(\ell)} \right|^2
	J(|\nu + \ell\omega |) 
\label{eq:SU1}
\end{equation}
does not depend on the Bose occupation numbers~(\ref{eq:BOP}) and
(\ref{eq:BOM}), while the second contribution does not depend on~$r$,
\begin{equation}
	R_2 = -c \sum_\ell	
	\hbar (\nu + \ell\omega) \, \left| v^{(\ell)} \right|^2
	N(\nu+\ell\omega) \, J(|\nu + \ell\omega |) \; . 
\label{eq:SU2}
\end{equation}
In the Appendix~\ref{App:A} we provide a formal proof of the intuitively 
expected, but non-obvious fact that this steady-state dissipation 
rate~(\ref{eq:DIS}) is positive, so that the energy flow always is directed 
from the driven system into the bath, regardless of the system's 
quasitemperature. In Fig.~\ref{F_5} we plot the dimensionless rate $R/R_0$, 
where the reference rate is taken as
\begin{equation}
	R_0 = \hbar\omega c J_0	
	\sum_\ell \left| v^{(\ell)} \right|^2 \; ,
\end{equation}	
for the situations previously considered in Figs.~\ref{F_3} and~\ref{F_4}. 
Evidently the total dissipation rate is duly positive, and diverges at the 
borders of quasithermal stability, as predicted by the prefactor of the
sum~(\ref{eq:SU1}).

\section{Discussion: Quasithermal engineering}
\label{sec:5}

The numerical examples worked out in the preceding section all rely on 
the proposition that the bath-specific input determining the partial 
rates~(\ref{eq:PRA}) and, hence, the quasistationary Floquet-state 
distributions $\{ p_n \}_{n = 0,1,2,\ldots}$, namely, the spectral density
$J(\wo)$ be given by the models~(\ref{eq:OHM}). With a view towards future 
applications of periodic thermo\-dynamics this assumption may not be realistic; 
a given system may interact with its environment preferentially at certain 
distinguished frequencies. As will be demonstrated now, spectral densities 
structured in this manner may have remarkable physical effects. Consider, for 
instance, a Gaussian density    
\begin{equation}
	J(\widetilde\omega) = 
	J_0 \exp\!\left(-\frac{(\wo - \wo_0)^2}{(\Delta\wo)^2}\right) \; .
\label{eq:GSD}
\end{equation}
Given a sufficiently narrow width $\Delta\wo$, and a central frequency~$\wo_0$ 
detuned not too far from one of the system's positive ``upward'' transition 
frequencies $\omega^{(\ell_1)}_{n+1,n} = \nu + \ell_1\omega$, this 
density~(\ref{eq:GSD}) will essentially reduce the numerator and the 
denominator of the ratio~(\ref{eq:EXR}) to the single contribution 
$\ell = \ell_1$, provided the accompanying squared Fourier coefficient is not 
too small, that is, if the drive is sufficiently strong. Since the transition 
frequencies enter into the density with their absolute value only, 
$J(|\nu + \ell_1\omega|)$ then cancels out of the remaining ratio, leaving us 
with  
\begin{equation}
	r \approx \frac{N(+\nu+\ell_1\omega)}{N(-\nu-\ell_1\omega)} \; .
\label{eq:CBD}
\end{equation}
Now the Bose occupation number $N(+\nu+\ell_1\omega) \equiv N_+$ is given 
by Eq.~(\ref{eq:BOP}), whereas $N(-\nu-\ell_1\omega) = N_+ + 1$ is obtained 
from Eq.~(\ref{eq:BOM}). If then additionally $N_+ \ll 1$, one deduces 
$r \approx N_+/(N_+ + 1) \ll 1$ --- meaning that the ``downward'' transitions 
can be strongly favored over the upward ones, even to the extent that the 
Floquet state $n = 0$ is populated with higher probability than the oscillator 
ground state in the absence of the drive, or, expressed differently, that the 
quasitemperature of the {\em driven\/} system is {\em lower\/} than the 
temperature of the bath it is coupled to~\cite{DiermannHolthaus19}.

\begin{figure}[t]
\centering
\includegraphics[width=0.9\linewidth]{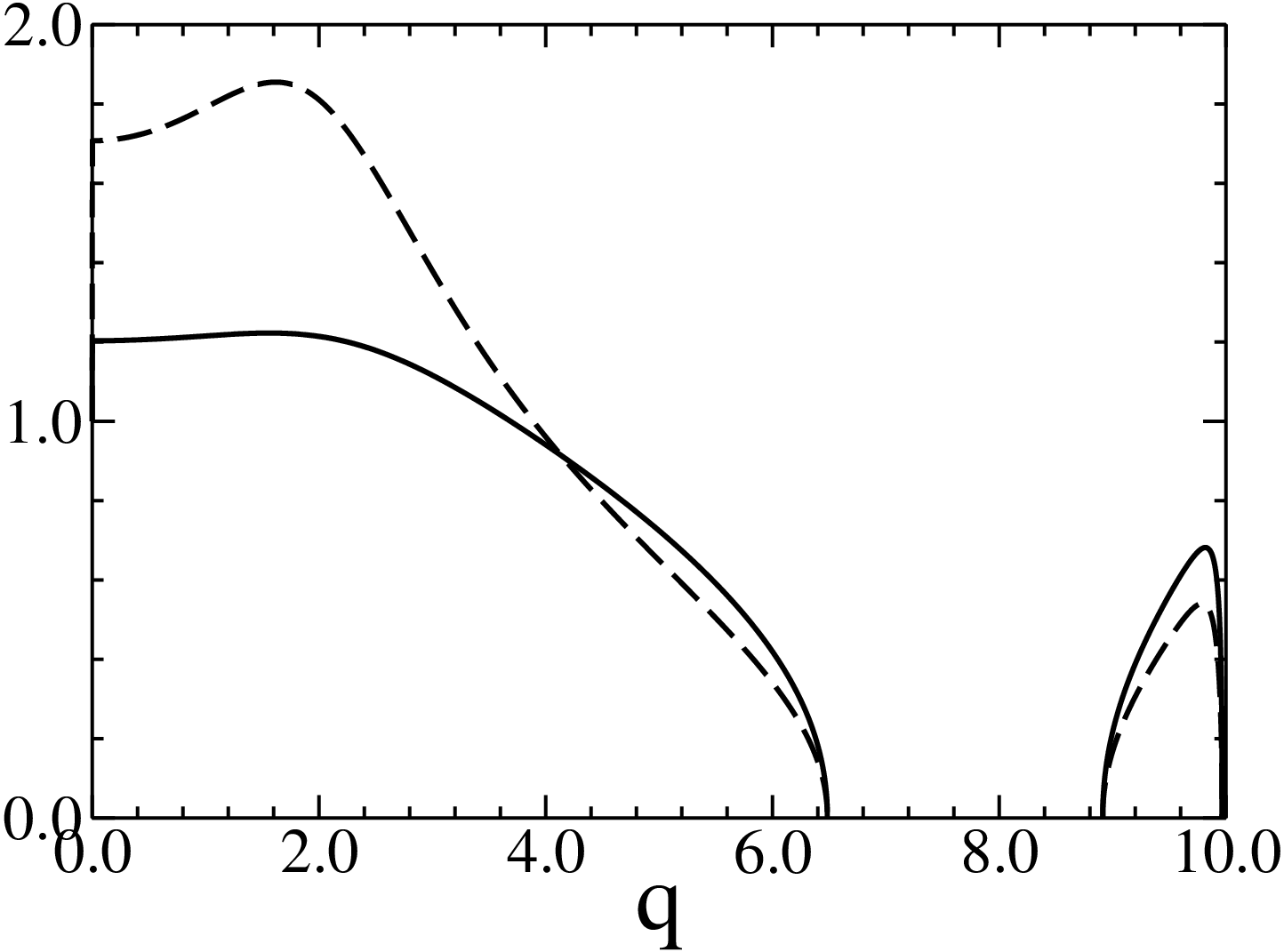}
\caption{Scaled inverse quasitemperature $\hbar\omega/(\kB\tau)$ (dashed) 
	and occupation probability $p_0/P_0$ (full line) for the Mathieu 
	oscillator with parameter $a = 8.0$ coupled to a heat bath with
	$\beta\hbar\omega = 1.0$ as in Fig.~\ref{F_3}, but with a Gaussian 
	spectral density~(\ref{eq:GSD}) centered around $\wo_0/\omega = 3.2$,
	with squared width $(\Delta\wo/\omega)^2 = 0.1$. Observe that the 
	quasitemperature of the driven system is {\em lower\/} than the bath 
	temperature for $0 < q \lesssim 3.89$. (Both lines bend sharply at 
	small~$q$ unresolved here, and connect to the ordinate~$1.0$ for 
	$q = 0$; see Fig.~\ref{F_7}.)} 
\label{F_6}
\end{figure}

To provide a working example of this counterintuitive ``cooling by driving'' 
mechanism, let us fix both the Mathieu parameter $a = 8.0$ and the scaled bath 
temperature $\beta\hbar\omega = 1.0$ to the values employed before, and let us 
select the parameters $(\Delta\wo/\omega)^2 = 0.1$ and $\wo_0/\omega = 3.2$ 
for the above density~(\ref{eq:GSD}). In view of Fig.~\ref{F_1}, showing that 
the canonical representative of the characteristic exponent then varies in the 
interval $\sqrt{2} \ge \nu/\omega \ge 1.0$ within the first zone of stability, 
this selection tends to favor the contributions with $\ell = \ell_1 = +2$ to 
the ratio~(\ref{eq:EXR}), but to an extent depending on the scaled driving 
strength because of the ac Stark shift of~$\nu$ with~$q$. Numerical data 
corresponding to this scenario are displayed in Fig~\ref{F_6}. Indeed, for 
$0 < q \lesssim 3.89$ one finds $\hbar\omega/(\kB\tau) > 1$, implying 
$\tau < T_{\rm bath}$: The driven system effectively is cooled.

\begin{figure}[t]
\centering
\includegraphics[width=0.9\linewidth]{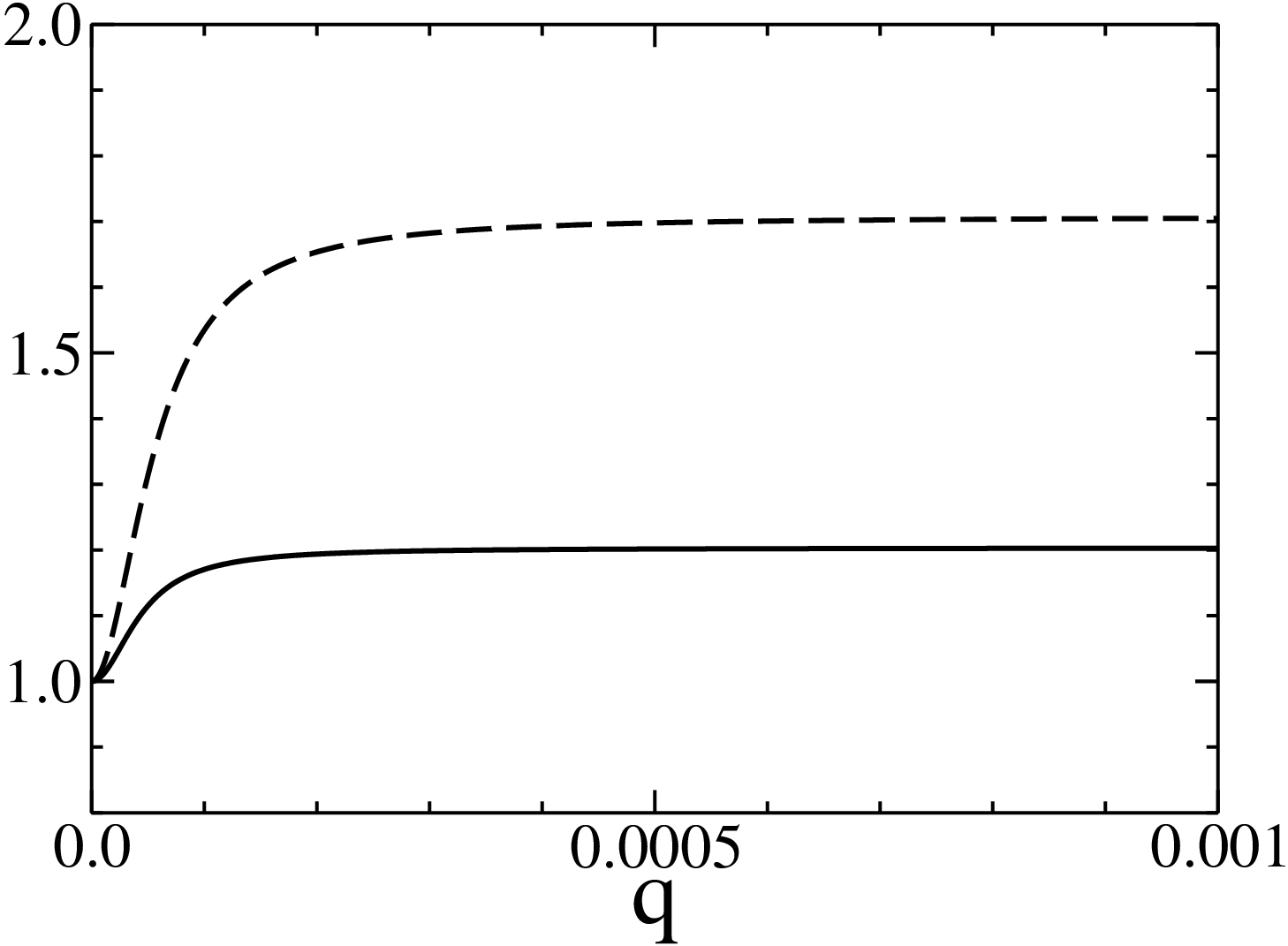}
\caption{As Fig.~\ref{F_6}, for small scaled driving strengths. The almost 
	constant values of $\hbar\omega/(\kB\tau)$ (dashed) and
	$p_0/P_0$ (full line) attained here are determined by 
	Eq.~(\ref{eq:CBD}) with $\ell_1 = 1$.}	
\label{F_7}
\end{figure}

Seemingly, the lines drawn in Fig.~\ref{F_6} do not connect to the ordinate
$1.0$ for vanishing~$q$, as they should. But actually, they do: In 
Fig.~\ref{F_7} we magnify the behavior of both $\hbar\omega/(\kB\tau)$ and
$p_0/P_0$ for very small $q$, confirming the expected continuity for $q \to 0$.
The plateau values adopted here are perfectly explained by Eq.~(\ref{eq:CBD}) 
with $\ell_1 = 1$, as the higher Fourier coefficients are still too small to 
yield sizable contributions. This case study indicates that ``cooling by 
driving'' may work even with fairly low driving strengths, although the 
corresponding relaxation times to the quasithermal nonequilibrium steady state 
may be quite long if the rates are small.

\begin{figure}[t]
\centering
\includegraphics[width=0.9\linewidth]{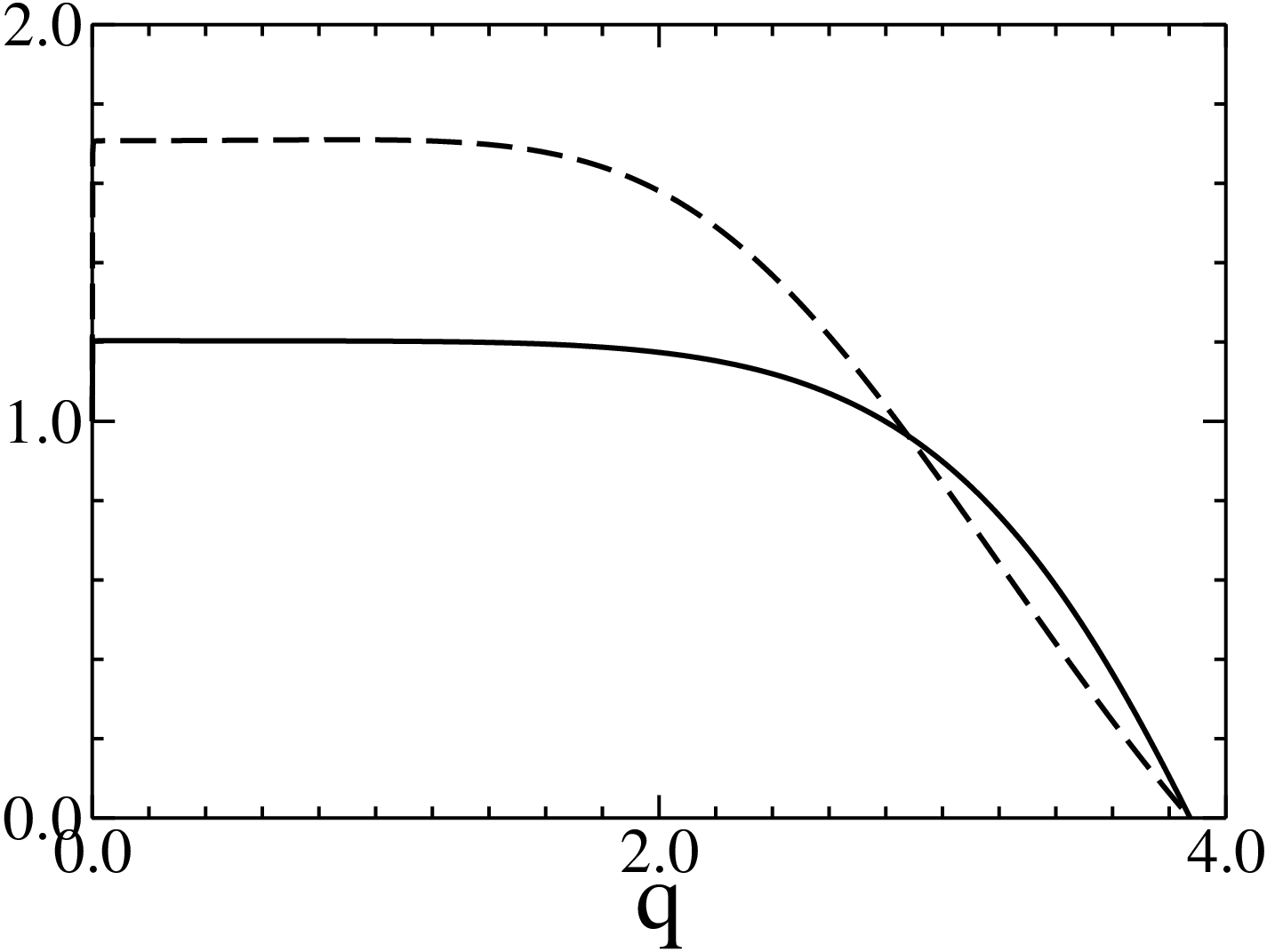}
\caption{As Fig.~\ref{F_6}, but with the Gaussian density~(\ref{eq:GSD}) now 
	being centered around $\wo_0/\omega = 3.0$. All other parameters have 
	remained unchanged. Observe that the onset of quasithermal instability
	here has been decoupled from the mechanical instability identified in 
	Fig.~\ref{F_1}, occurring already at $q \approx 3.87$.}
\label{F_8}
\end{figure}

The possibilities to shape a quasistationary Floquet-state distribution with 
the help of the spectral density of the bath are by no means exhausted by this 
inaugural example of ``quasithermal engineering''. Shifting the peak position 
from $\wo_0/\omega = 3.2$ to $\wo_0/\omega = 3.0$, but leaving all other 
parameters at their values already used for Fig.~\ref{F_6}, one obtains the 
data visualized in Fig.~\ref{F_8}: Here the onset of thermal instability 
already occurs for significantly lower driving strength $q \approx 3.87$ than 
the mechanical instability spotted in Fig.~\ref{F_1} at $q \approx 6.49$, and
the system does not become quasithermally stable in the second regime of
mechanical stability. This is no contradiction to our previous finding that
$r \to 1$ at a mechanical stability border, since $r$ raises to values higher
than $1.0$ already at $q \approx 3.87$, and then approaches unity from above.  

Thus, the fact that quasistationary Floquet-state distributions do depend 
on the precise form of the system-bath coupling~\cite{Kohn01,BreuerEtAl00} 
allows one to achieve unexpected effects by deliberately designing this 
coupling, that is, by quasithermal engineering. The phenomenon of ``cooling 
by driving'', which reflects one particular application of this concept, 
bears interesting promises: If it were possible to decouple a driven system 
with a quasitemperature~$\tau$ lower than the temperature $T_{\rm bath}$ 
of its bath from that bath, and then switch off the drive in an adiabatic 
manner so that the Floquet-state occupation probabilities would be preserved, 
the system would end up in state with a genuine temperature 
$\tau < T_{\rm bath}$~\cite{LangemeyerHolthaus14}.    

The model system we have employed in the present study, the parametrically 
driven Mathieu oscillator, still is exceptionally simple from the Floquet 
point of view, not showing features which are characteristic for more 
generic non-integrable systems~\cite{KetzmerickWustmann10,HoneEtAl09}. 
When dealing with such generic systems, one has to compute the Floquet 
states fully numerically in order to obtain the transition matrix
elements~(\ref{eq:TME}) and their Fourier components, and then requires
a numerical solution of the master equation~(\ref{eq:PME}), thus obstructing
a clear view on the underlying physics. This is why simplicity is an 
outstanding virtue here. The only ``hard'' data required for converting 
predictions made by our model into numbers are the Fourier coefficients of 
the periodic parts~(\ref{eq:FSV}) of the solutions to the classical equation 
of motion~(\ref{eq:EOM}); even these coefficients can be obtained with fairly 
modest numerical effort. Therefore, the parametrically driven harmonic 
oscillator coupled to a thermal bath in various manners may serve as a 
valuable source of inspiration in the further exploration of periodic 
thermodynamics.

\begin{acknowledgments}
This work has been supported by the Deutsche For\-schungsgemeinschaft (DFG,
German Research Foundation) through Project No.~397122187. One of us (M.H.) 
wishes to thank the members of the Research Unit FOR~2692 for insightful 
discussions. In particular, he is indebted to Heinz-J\"urgen Schmidt for 
instructive comments concerning the conditions for the emergence of detailed 
balance.
\end{acknowledgments}

\begin{appendix}
\section{Positivity of the dissipation rate}
\label{App:A}

In this Appendix we demonstrate that the dissipation rate~(\ref{eq:DIS}) 
is always positive, which implies that the energy flows from the driven 
oscillator into the bath when the system is in a quasistationary state,
regardless of whether the system's quasitemperature is higher or lower than 
the temperature of the bath it is coupled to. Accounting for $m = n \pm 1$ 
and $\omega_{n\pm1,n}^{(\ell)} = \pm\nu + \ell\omega$, one has
\begin{eqnarray}
	R & = & -\sum_{n\ell} 
	\hbar(+\nu + \ell\omega) \, \Gamma_{n+1,n}^{(\ell)} \; p_n
\nonumber \\	
	& & -\sum_{n\ell}
	\hbar(-\nu + \ell\omega) \, \Gamma_{n-1,n}^{(\ell)} \; p_n \; . 	
\label{eq:SRT}
\end{eqnarray}	
The proof of the positivity of~$R$ rests on the observation that the
contribution to this expression which is proportional to $\nu$ vanishes: 
We find
\begin{eqnarray}
	& &
	-\hbar\nu \sum_{n\ell} \Big( 
	\Gamma_{n+1,n}^{(\ell)} - \Gamma_{n-1,n}^{(\ell)} \Big) p_n
\nonumber \\	& = &
	-\hbar\nu \sum_{n} \Big( 
	\Gamma_{n+1,n} \; p_n - \Gamma_{n,n+1} \; p_{n+1} \Big) 
\nonumber \\	& = & 0 \; ,	 	 		
\end{eqnarray}
where, successively, the definition~(\ref{eq:GFI}) and the detailed-balance
relation~(\ref{eq:FRE}) have been exploited. This allows us to replace $\nu$ 
in Eq.~(\ref{eq:SRT}) by $\nu + \chi$, with arbitrary~$\chi$. Therefore, 
the previous representation~(\ref{eq:R12}) involving the two 
expressions~(\ref{eq:SU1}) and~(\ref{eq:SU2}) can be cast into the form    
\begin{widetext}
\begin{eqnarray}
	R & = & \phantom{+}
	\hbar c \sum_{\nu + \ell\omega > 0}	
	\big(\nu + \ell\omega + \chi \big) \, \left| v^{(\ell)} \right|^2
	J(|\nu + \ell\omega |)
	\left( \frac{r}{1-r} - N(\nu+\ell\omega) \right)
\nonumber \\	& &
	+ \hbar c \sum_{\nu + \ell\omega < 0}
	\big(|\nu + \ell\omega| - \chi \big) \, \left| v^{(\ell)} \right|^2
	J(|\nu + \ell\omega |)
	\left( \frac{r}{1-r} + N(\nu+\ell\omega) \right) \; .		 
\label{eq:SEP}
\end{eqnarray}
\end{widetext}
Now consider the last factor of the contributions to the first sum, namely,
\begin{eqnarray}
	& &
	\frac{r}{1-r} - N(\nu+\ell\omega) 
\nonumber \\	& = &
	\frac{r\,\exp\!\big(\beta\hbar[\nu+\ell\omega]\big) - 1}
	{(1-r)(\exp\!\big(\beta\hbar[\nu+\ell\omega]\big) - 1)} \; .
\end{eqnarray}
Since $\nu + \ell\omega$ is positive here, both factors appearing in the
denominator on the right-hand side are positive for a quasistationary state
with $0 < r < 1$, while the numerator, which increases monotonically 
with~$\ell$, may change its sign, being negative for $\ell < \ell_0$ and
non-negative for $\ell \ge \ell_0$. Let us then select $\chi \le 0$ such that
the factor $\nu + \ell\omega + \chi$ is negative for $\ell < \ell_0$, but
positive for $\ell \ge \ell_0$. Then all terms contributing to the first
sum in Eq.~(\ref{eq:SEP}) are non-negative, while the second sum is manifestly 
positive for $\chi \le 0$. This concludes the proof.  

\end{appendix}

\end{document}